\documentclass[12pt]{article}
\usepackage{fullpage}
\usepackage{cite}
\usepackage{amstext,amsfonts,amsbsy,eucal,amssymb}
\usepackage{epsfig}
\usepackage{psfrag}

\newcounter{definition}

\def\thefigure{\arabic{section}.\arabic{figure}}
\def\theequation{\thesection.\arabic{equation}}
\def\appendix{
  \setcounter{section}{0}
  \setcounter{subsection}{0}
  \par
  \def\thesection{Appendix \Alph{section}}
  \def\theequation{\Alph{section}.\arabic{equation}}
  \def\thefigure{\Alph{section}.\arabic{figure}}
}
\makeatletter
\def\fnum@figure{Fig. \thefigure}
\@addtoreset{equation}{section}
\@addtoreset{figure}{section}
\makeatother

\date{
\today
}
\title{
Third-neighbor and other four-point correlation
functions of spin-1/2 XXZ chain
} 
\author{
\small
  Go {\sc Kato}${}^1$\thanks{kato@monet.phys.s.u-tokyo.ac.jp},
  Masahiro {\sc Shiroishi}${}^2$\thanks{siroisi@issp.u-tokyo.ac.jp},
  Minoru {\sc Takahashi}${}^2$\thanks{mtaka@issp.u-tokyo.ac.jp}\, and
  Kazumitsu {\sc Sakai}${}^3$\thanks{sakai@stat.phys.titech.ac.jp}
 \\\it
  ${}^1$Department of Physics, Graduate School of Science,
  University of Tokyo,\\\it
  Hongo 7-3-1, Bunkyo-ku, Tokyo 113-0033, Japan\\\it
  ${}^2$Institute for Solid State Physics, University of Tokyo,\\\it 
  Kashiwanoha, Kashiwa-shi, Chiba 277-8571, Japan\\\it
  ${}^3$Department of Physics, Tokyo Institute of Technology,\\\it
  Oh-okayama, Meguro-ku, Tokyo 152-8551, Japan
}

\begin{document}
\maketitle
\setlength{\baselineskip}{1.8em}
\begin{abstract}
The correlation functions of the spin-1/2 $XXZ$ chain in the ground state were expressed in the
 form of multiple integrals for $-1<\Delta\leq1$ and  $1<\Delta$.
In particular, adjacent four-point correlation functions were given as
 certain four-dimensional integrals.
 We show that these integrals can be
 reduced to polynomials with respect to specific
 one-dimensional integrals. The results give the polynomial
 representation of the third-neighbor correlation functions.
\end{abstract}

\newpage
\section{Introduction}
The one-dimensional spin-1/2 $XXZ$ model is a quantum integrable
system and has been well studied in statistical physics.
The Hamiltonian is given by 
\begin{eqnarray}
H=\sum_{j=-\infty}^{\infty}\left[S^x_jS^x_{j+1}+S^y_jS^y_{j+1}+\Delta S^z_jS^z_{j+1}\right].
 \label{eq:XXZ_Hamiltonian}
\end{eqnarray}
where $S^\alpha=\sigma^\alpha/2 (\alpha=x,y,z)$ with $\sigma^\alpha$ being Pauli matrices. Here 
${\Delta}$ is the anisotropy parameter. 
The eigenvalues and the eigenfunctions of the system were obtained by Bethe ansatz~\cite{Bethe31,
KorepinBook}, which allowed us to investigate many physical quantities~\cite{TakahashiBook}. 
However, as for the correlation functions, it is not fully understood yet except for the 
${\Delta=0}$ case, where the system reduces to the free-fermion model~\cite{Lieb61,McCoy68,
Shiroishi01,Kitanine02c}. 

For general ${\Delta \ne 0}$ cases, nevertheless, there are also some recent progresses in 
the evaluation of the correlation functions, especially for the spin-spin correlation functions (SSCF) 
$\left<S_0^\alpha S_m^\alpha\right>$ and the emptiness formation probability (EFP) 
${\left< \prod_{j=1}^m(S_j^z+{\frac12}) \right>}$~\cite{Korepin94}. For example, via the field 
theoretical approach, we can now discuss the long-distance asymptotic behavior of the 
SSCF~\cite{Luther75,Lukyanov97,Lukyanov99,Lukyanov03}, which decays algebraicly in the critical region 
${-1<\Delta \le 1}$. The peculiar Gaussian decay of the EFP is also studied recently~\cite{Kitanine02d, 
Korepin03}.  However these works contain some assumptions or approximations. Hence we usually 
need some other means, such as numerical simulations, to appeal their validity. So it is more desirable 
if we can evaluate the correlations for an arbitrary ${m}$ exactly first, and derive its asymptotic behavior 
thereafter.\footnote{Note that such a program has been achieved for the asymptotic behavior of the EFP 
${P(m)}$ at ${\Delta=1/2}$.}  

In fact it has been known already that the analytical expressions for the SSCF and the EFP at arbitrary 
$m$ as well as other correlations among adjacent ${m}$-sites are given by certain ${m}$-dimensional 
integrals~\cite{Jimbo92,Jimbo95,Nakayashiki94,Jimbo96,Kitanine00, Kitanine02a, Kitanine02b}. Unfortunately 
there is a serious problem for these multiple integrals that it is hard to evaluate them, even numerically, 
because of the complicated structures of the integrand.  For this reason, it has been an important and 
challenging problem to simplify these multiple integral expressions systematically. Actually, a few years ago, 
Boos {\it et al} succeeded in simplifying the multiple integrals for EFP at $\Delta=1$. They could  express 
the EFP up to ${m=6}$ as polynomials with respect to the specific values of $\zeta$ function 
\cite{Boos01,Boos02,BKNS02,Boos03}. The same method was 
applied to the third-neighbor correlations at $\Delta=1$\cite{Sakai03}.  Furthermore, we have recently 
succeeded in generalizing the method to the general anisotropy parameter 
$\Delta$. Particularly we have reported in our previous papers, 
that the three-dimensional integrals relevant for the second 
neighbor correlations can be simplified to one-dimensional 
ones \cite{kato_corelation_a,kato_corelation_b}. The main purpose of this paper is to present 
a detailed account of our method. Further, applying our 
method, we evaluate all the adjacent four point correlations 
including the third-neighbor correlation functions  for general ${\Delta}$. Our significant 
finding is that these correlation functions are expressed as 
polynomials with respect to the following specific one-dimensional 
integrals,
\begin{eqnarray}
\zeta _\eta\left(j\right)&:=&\int_{C_-}dx\frac1{\sinh x}\frac{\cosh \eta x}{ \sinh^j \eta x},
\label{ea:notation_first_intro}\\
\zeta'_\eta\left(j\right)&:=&\int_{C_-}dx\frac1{\sinh x}\frac{\partial}{\partial\eta}\frac{\cosh \eta x}{ \sinh^j \eta x},
\label{ea:notation_last_intro}
\end{eqnarray}
where the integration path is given by $C_-=(-\infty-\pi i/2,\infty-\pi i/2)$ and the anisotropy parameter 
$\eta$ ( ${\Delta = \cos \pi \eta}$) is either a real number from $0$ to $1$ or  a purely imaginary number. 
Probably this property will be valid for any correlation functions 
for the ${XXZ}$ chain (cf. \cite{Boos03b}). 
We also remark the one-dimensional integrals (\ref{ea:notation_first_intro}) and (\ref{ea:notation_last_intro}) can be 
integrated analytically when $\eta$ takes a real rational number. In such a
case, we can obtain the complete analytic values for the correlation 
functions.

The outline of this paper is the following. In \S2,
 we show a general strategy to evaluate the multiple integral at the massless
 case. In \S3, we point out the similarity of the procedures in the massless case and 
the massive case. In \S4, we evaluate, as an example, an adjacent two-point
correlation function by the strategy explained in \S2 and \S3 . This derivation does not 
depend on $\Delta$, i.e. it works both in the massive and the massless cases.
In \S5, we show the polynomial representation for the third-neighbor correlation functions 
and some of their analytical values at special points of ${\Delta}$.
 The last section is devoted to the concluding remarks.
The outline of the derivation for an adjacent four-point correlation function is presented in Appendix A.
In Appendix B, we list the polynomial expressions of all the independent  
 correlation functions among up to four adjacent points.

\section{General discussion at the massless case}
In the massless region ${-1<\Delta \le 1}$ the correlation functions 
$F\left[^{\epsilon'_1\,\cdots\,\epsilon'_n}_{\epsilon_1\,\cdots\, \epsilon_n}\right]$
among adjacent ${n}$-sites are given by the following multiple integrals~\cite{Jimbo96}, 
\begin{eqnarray}
&&F\left[
\begin{array}{ccc}
\epsilon'_1 &\cdots &\epsilon'_n\\  
\epsilon _1 &\cdots &\epsilon _n
\end{array}
\right]
\: := \:
 \left<
E^{\left(1\right)}_{\epsilon'_1 \epsilon_1}
\cdots
E^{\left(n\right)}_{\epsilon'_n \epsilon_n}
\right>
\nonumber\\
&&=
\nu^{-n\left(n-1\right)/2}
\prod_{a\in A_+}
\int_{C_+}\frac{-dx_a}{2\pi i}
\prod_{a\in A_-}
\int_{C_-}\frac{dx_a}{2\pi i}
\prod_{1\leq l<l'\leq n}\frac{\sinh\left(x_{l}-x_{l'}\right)}{\sinh\nu\left(x_{l}-x_{l'}-\pi i\right)}
\nonumber\\&&{}
\prod_{a\in A_+}
\frac
{
 \sinh^{\bar a-1}\nu x_a
 \cdot
\sinh^{n-\bar a}\nu \left(x_a-\pi i\right)
}
{\sinh^n x_a}
\prod_{a\in A_-}
\frac
{
 \sinh^{\bar a-1}\nu x_a
 \cdot
\sinh^{n-\bar a}\nu \left(x_a+\pi i\right)
}
{\sinh^n x_a}.
\label{eq:integral_formula_P_n}
\end{eqnarray}
Here $\epsilon$ and
$\epsilon'$ take either $+$ or $-$, and $E_{\epsilon' \epsilon}$ are the $2\times 2$ matrix with 1 at
the $(\epsilon',\epsilon)$-element and 0 elsewhere. $E^{(j)}_{\epsilon' \epsilon}$ denote the 
operators acting as $E_{\epsilon' \epsilon}$ on the $j$-th spin site and as the identity on the other 
sites. Then the correlation function $\left<E^{\left(1\right)}_{\epsilon'_1 \epsilon_1}
\cdots
E^{\left(1\right)}_{\epsilon'_n \epsilon_n}
\right>$ is defined by the expected value of the operator 
$E^{\left(1\right)}_{\epsilon'_1 \epsilon_1} \cdots
E^{\left(n\right)}_{\epsilon'_n \epsilon_n}
$ with respect to
the ground state of the $XXZ$ Hamiltonian (\ref{eq:XXZ_Hamiltonian}).
For the description of the multiple integrals, we use the same notations as  
\cite{Jimbo96}. Let us note them briefly below. 
First the anisotropy parameter $\nu$ is defined by the relations 
$\Delta=\cos \pi\nu $. As we are considering the massless region ${-1<\Delta\le 1}$ here, 
the parameter takes $ 0\leq\nu<1$. The contours $C_{\pm}$ in each integral goes from 
$\pm\pi i/2-\infty$ to $\pm\pi i/2 +\infty$.
The bars, e.g. $ \bar a$, denotes the mapping 
from $\{1,2,\cdots,n\}$ to $\{1,2,\cdots,n\}$ defined by the following
condition.
The $+$'s in the sequence
 $-\epsilon'_1,\cdots, -\epsilon'_n,\epsilon_n,\cdots ,\epsilon_1$ are
 $-\epsilon'_{\bar 1},\cdots, -\epsilon'_{\bar s},\epsilon_{\overline{s+1}},\cdots ,\epsilon_{\bar n}$ where $s$ is the number of $-$ in the
 sequence $\epsilon'_1,\cdots,\epsilon'_n$.
The sets $A_+$ and $A_-$ are defined as $\{j|\epsilon'_{\bar j}=-\}$ and
$\{j|\epsilon_{\bar j}=+\}$ respectively. Note that here we assume ${\#A_+ + \#A_- = n}$, 
because otherwise the expected value $F\left[^{\epsilon'_1\,\cdots\,\epsilon'_n}_{\epsilon_1\,\cdots\,
 \epsilon_n}\right]$ is equal to zero.

Let us describe a strategy that may be used for the
evaluation of general
 $F\left[^{\epsilon'_1\,\cdots\,\epsilon'_n}_{\epsilon_1\,\cdots\,
 \epsilon_n}\right]$.
We first represent the integral formula (\ref{eq:integral_formula_P_n}) as follows:
\begin{eqnarray}
&&
F\left[
\begin{array}{ccc}
\epsilon'_1 &\cdots &\epsilon'_n\\  
\epsilon _1 &\cdots &\epsilon _n
\end{array}
\right]
\nonumber\\
&&= 
\nu^{-n\left(n-1\right)/2}
\prod_{j=1}^n
\int_{C_-}\frac{dx_j}{2\pi i}
U\left(x_1,\cdots,x_n\right)
T\left(\exp(2x_1 \nu),\cdots,\exp(2x_n \nu)\right),
\label{eq:simplified_integral_form_P_n}
\end{eqnarray}
where
\begin{eqnarray}
U\left(x_1,\cdots,x_n\right)
&=&
\frac
{
\prod_{1\leq l<l'\leq n}
\sinh\left(x_l-x_{l'}\right)
}
{\prod_{l=1}^n\sinh^n x_l},
\label{eq:definition_U}
\end{eqnarray}
\begin{eqnarray}
&&T( q^{-1}y_1,\cdots, q^{-1}y_s,y_{s+1},\cdots,y_n)
\nonumber\\
&&=
\frac
{
\prod_{a=A_+}
\left(y_a-1\right)^{\bar a-1}
\left(q^{-\frac12} y_a - q^{\frac12}\right)^{n-\bar a}
\prod_{a=A_-}
\left(y_a-1\right)^{\bar a-1}
\left( q^{\frac12}  y_a- q^{-\frac12}\right)^{n-\bar a}
}
{2^{n(n-1)/2}
\prod_{1\leq l,l'\leq n}
\left(q^{-\frac12} y_l -q^{\frac12} y_{l'} \right)
},
\label{eq:define_T}
\end{eqnarray}
and
\begin{eqnarray}
 q&\equiv&\exp\left( 2\pi i \nu\right).
\end{eqnarray}
We can make a lot of simplification without evaluating 
integration but using only some simple observations and properties of the 
integrand of (\ref{eq:simplified_integral_form_P_n}). For this purpose, let us first 
define a ``weak'' equality $\sim$. We call that two
functions $F(y_1,\cdots,y_n)$ and $G(y_1,\cdots,y_n)$ are ``weakly''
equivalent 
\begin{eqnarray}
 F\left(y_1,\cdots,y_n\right)&\sim&G\left(y_1,\cdots,y_n\right),
\end{eqnarray}
if
\begin{eqnarray}
\prod_{j=1}^n
\int_{C_-}\frac{dx_j}{2\pi i}
U\left(x_1,\cdots,x_n\right)
\left[
 F\left(e^{2x_1 \nu},\cdots,e^{2x_n \nu}\right)
-G\left(e^{2x_1 \nu},\cdots,e^{2x_n \nu}\right)
\right]
&=&0.
\end{eqnarray}

Let us also introduce a ``canonical'' form of the function by the
following formula
\begin{eqnarray}
 T^c\left(y_1,\cdots,y_n\right)
&=&
\sum_{j=0}^{\left[\frac n2\right]}
P_j^{(n)}\prod_{k=1}^j\frac1{y_{2k}-y_{2k-1}},
\label{eq:define_T^c}
\end{eqnarray}
where $P_j^{(n)}$ are some polynomials with respect to $y_k$ such that
\begin{eqnarray}
 P_j^{(n)}&\equiv&P_j^{(n)}\left(y_1,y_3,\cdots,y_{2j-1}|y_{2j+1},y_{2j+2},\cdots,y_n\right)\nonumber\\
&=&
\sum_{
\begin{array}{c}

 0\leq i_1\leq i_3\leq\cdots \leq i_{2j-1}\leq n-1
\\ 
 0\leq i_{2j+1} <i_{2j+2} < \cdots < i_n\leq n-1
\end{array}
} C_{i_1,i_3,\cdots,i_{2j-1}}^{i_{2j+1},i_{2j+2},\cdots,i_n}y_1^{i_1}\cdots y_n^{i_n}
\label{eq:define_P_j}
\end{eqnarray}
and the coefficients $C_{i_1,\cdots,i_{2j-1}}^{i_{2j+1},\cdots,i_n}$ are
rational functions of $q^{1/2}$.
Our hypothesis is that for any $n$ and $\epsilon_m$ one can reduce the function $T$
defined by (\ref{eq:define_T}) to the canonical form, i.e. there exist
polynomials in (\ref{eq:define_T^c})
such that
\begin{eqnarray}
 T\left(y_1,\cdots,y_n\right)&\sim&T^c\left(y_1,\cdots,y_n\right).
\label{eq:general_integrant}
\end{eqnarray}
Unfortunately, for the moment we do not have a proof of this statement
for any $n$ but we demonstrate how it works for $n=2,4$  in the following sections.

In fact, the problem of the evaluation of
 $F\left[^{\epsilon'_1\,\cdots\,\epsilon'_n}_{\epsilon_1\,\cdots\,
 \epsilon_n}\right]$
 given by the integral (\ref{eq:simplified_integral_form_P_n}) can be reduced to two steps. 
The first step corresponds to the
transformation into the ``canonical'' form.
The second step is the integration by means of this
 ``canonical'' form. For these procedures, we need several relations as follows.
\begin{itemize}
 \item {\bf Proposition 1}

Since the function $U\left(x_1,\cdots,x_n\right)$ is
       antisymmetric with respect to transposition of any pair of
       integration variables $x_j$ and $x_k$, we have 
\begin{eqnarray}
\prod_{j=1}^n
\int_{C_-}\frac{dx_j}{2\pi i}
U\left(x_1,\cdots,x_n\right)
S\left(\exp(2x_1 \nu),\cdots,\exp(2x_n \nu)\right)&=&0,
\label{eq:proposition_1}
\end{eqnarray}
if the function $S$ is symmetric for at least one pair of
       $x_s$. Therefore for an arbitrary function $f(y_1,\cdots,y_n)$,
       we can transpose any pair of $y_s$ taking into consideration
       appearance of additional sign due to the antisymmetry of
       $U(x_1,\cdots,x_n)$. For example, by transposing $y_j$ with
       $y_k$, we get
\begin{eqnarray}
 f\left(y_1,\cdots,y_j,\cdots,y_k,\cdots,y_n\right)
&\sim&
-f\left(y_1,\cdots,y_k,\cdots,y_j,\cdots,y_n\right).
\label{eq:transposition}
\end{eqnarray}

\item {\bf Proposition 2}

Let the rational function $f\left(y_1,\cdots,y_n\right)$
have poles of the terms only $1/(y_j- q^a y_k  )$ and $1/y_j^a$
where $a$ is an integer,
i.e. the product $U(x_1,\cdots,x_n)f(e^{2x_1\nu},\cdots,e^{2x_n\nu})$
does not have poles of the terms $1/(e^{2x_j\nu}-q^a e^{2x_k\nu}  )$. Then
\begin{eqnarray}
 \left(y_j-1\right)^mf\left(y_1,\cdots,y_j,\cdots,y_n\right)
&\sim&
 -\left(q y_j -1\right)^mf\left(y_1,\cdots,q y_j ,\cdots,y_n\right),
\label{eq:proposition1}
\end{eqnarray}
where $m$ is an integer and $m\geq n$.
\end{itemize}
We can get four useful corollaries from the proposition 1 and 2.
\begin{itemize}
\item {\it Corollary 1}
\begin{eqnarray}
&&\frac{y_k^a y_l^b}{\left(y_j-y_k\right)\left(y_j-y_l\right)}g\left(y_1,\cdots, \hat y_k,\cdots,\hat y_l,\cdots, y_n\right)
\nonumber\\&\sim&
\frac{\left(y_k y_l\right)^b\sum_{m=0}^{a-b-1} y_k^m y_l^{a-b-1-m}}{\left(y_j-y_k\right)}g\left(y_1,\cdots, \hat y_k,\cdots,\hat y_l,\cdots, y_n\right),
\label{eq:reduce_pole}
\end{eqnarray}
where $a>b$ and
the function $ g\left(y_1\cdots,\hat y_k,\cdots,\hat y_l,\cdots,y_n\right)$ does
       not depend on $y_k$ and $y_l$.
The relation (\ref{eq:reduce_pole}) also holds in the case $y_j$ is replaced
       with $q y_j$ or $q^{-1} y_j$. 

{\it proof}
\begin{eqnarray}
&&
\frac{\left(y_k y_l\right)^b\sum_{m=0}^{a-b-1} y_k^m y_l^{a-b-1-m}}{\left(y_j-y_k\right)}g\left(y_1,\cdots, \hat y_k,\cdots,\hat y_l,\cdots, y_n\right)
\nonumber\\&\sim&
\frac{\left(y_k y_l\right)^b\sum_{m=0}^{a-b-1} y_k^m y_l^{a-b-1-m}}{2\left(y_j-y_k\right)}g\left(y_1,\cdots, \hat y_k,\cdots,\hat y_l,\cdots, y_n\right)
\nonumber\\ &&{}-
\frac{\left(y_k y_l\right)^b\sum_{m=0}^{a-b-1} y_k^m y_l^{a-b-1-m}}{2\left(y_j-y_l\right)}g\left(y_1,\cdots, \hat y_k,\cdots,\hat y_l,\cdots, y_n\right)
\nonumber\\&=&
\frac{y_k^a y_l^b-y_l^a y_y^b}{2\left(y_j-y_k\right)\left(y_j-y_l\right)}g\left(y_1,\cdots, \hat y_k,\cdots,\hat y_l,\cdots, y_n\right)
\nonumber\\&\sim&
\frac{y_k^a y_l^b}{\left(y_j-y_k\right)\left(y_j-y_l\right)}g\left(y_1,\cdots, \hat y_k,\cdots,\hat y_l,\cdots, y_n\right).
\end{eqnarray}
At the two weak equalities above, we have used the proposition 1.

\item {\it Corollary 2}
\begin{eqnarray}
&&\frac{y_j^a y_k^by_l^cy_m^a}{\left(y_j-y_k\right)\left(y_k-y_l\right)\left(y_l-y_m\right)}g\left(y_1,\cdots, \hat y_j\cdots, \hat y_k,\cdots,\hat y_l,\cdots, \hat y_m\cdots,y_n\right)
\nonumber\\&\sim&
\frac{y_j^ay_m^a\left(y_k y_l\right)^c\sum_{m=0}^{b-c-1} y_k^m y_l^{b-c-1-m}}{2\left(y_j-y_k\right)\left(y_l-y_m\right)}g\left(y_1,\cdots, \hat y_j\cdots, \hat y_k,\cdots,\hat y_l,\cdots, \hat y_m\cdots,y_n\right),
\nonumber\\
\label{eq:reduce_pole_2}
\end{eqnarray}
where $b>c$ and
the function $ g\left(y_1,\cdots,\hat y_j\cdots,\hat y_k,\cdots,\hat y_l,\cdots,\hat y_m,\cdots,y_n\right)$ does
       not depend on $y_j$, $y_k$, $y_l$ and $y_m$.
The relation (\ref{eq:reduce_pole_2}) is proved as follows.
\begin{eqnarray}
&&\frac{y_j^a y_k^by_l^cy_m^a}{\left(y_j-y_k\right)\left(y_k-y_l\right)\left(y_l-y_m\right)}g\left(y_1,\cdots, \hat y_j\cdots, \hat y_k,\cdots,\hat y_l,\cdots, \hat y_m\cdots,y_n\right)
\nonumber\\&\sim&
\frac{y_j^a \left(y_k^by_l^c-y_l^by_k^c\right)y_m^a}{2\left(y_j-y_k\right)\left(y_k-y_l\right)\left(y_l-y_m\right)}g\left(y_1,\cdots, \hat y_j\cdots, \hat y_k,\cdots,\hat y_l,\cdots, \hat y_m\cdots,y_n\right)
\nonumber\\&=&
\frac{y_j^ay_m^a\left(y_k y_l\right)^c\sum_{m=0}^{b-c-1} y_k^m y_l^{b-c-1-m}}{2\left(y_j-y_k\right)\left(y_l-y_m\right)}g\left(y_1,\cdots, \hat y_j\cdots, \hat y_k,\cdots,\hat y_l,\cdots, \hat y_m\cdots,y_n\right).
\nonumber\\
\end{eqnarray}
Here we used the proposition 1 similarly at the weak equality.

\item {\it Corollary 3}
\begin{eqnarray}
 y_j^{l+p} g\left(y_1,\cdots,\hat y_j,\cdots,y_n\right)
&\sim&
-\sum_{k=0,\neq l}^{m}\frac{a_k}{a_l}y_j^{k+p}
g\left(y_1,\cdots,\hat y_j,\cdots,y_n\right),
\label{eq:corollary_1}
\end{eqnarray}
where
\begin{eqnarray}
 a_k
&=&
\left(-\right)^{k}\frac{m!}{k!\left(m-k\right)!}\left(1+ q^{k+p}\right),
\end{eqnarray}
$p,l$ are integers and the function $g\left(y_1,\cdots,\hat y_j,\cdots,y_n\right)$ does
       not depend on $y_j$ and as above it is implied that $m\geq n$.
In our calculation, corollary 3 is used only in the case $l=0$ or $m$.

{\it proof}

The relation (\ref{eq:corollary_1}) is derived easily from the relation 
$
((y_j-1)^m y_j^p+( q y_j-1)^m( q y_j)^p)$ 
$g\left(y_1,\cdots,\hat y_j,\cdots,y_n\right)
\sim 0$, i.e. the proposition 2. 

\item {\it Corollary 4}
\begin{eqnarray}
&&
\frac{ y_j^{s+p}}{y_k-y_j} g\left(y_1,\cdots,\hat y_k,\cdots,\hat y_j,\cdots,y_n\right)
\nonumber\\
&\sim&
\left(
-\sum_{l=0,\neq s}^{m}
\frac{b_l}{b_s} \frac{y_j^{l+p}}{y_k-y_j}
+\sum_{l=0}^{m}
\frac{c_l}{b_s}
\sum_{r=0}^{l+p-1}
y_k^r\left(qy_j\right)^{l+p-r-1}
\right)
 g\left(y_1\cdots,\hat y_k,\cdots,\hat y_j,\cdots,y_n\right),
\nonumber\\
\label{eq:corollary_2}
\end{eqnarray}
where 
\begin{eqnarray}
 b_l
\; = \;
\left(-\right)^{m-l}
\frac{
m!
}{l!\left(m-l\right)!
}\left(
1- q^{l+p-1}
\right),
\quad\quad
 c_l
\; = \;
\left(-\right)^{m-l}
\frac{
m!
}{l!\left(m-l\right)!
},
\end{eqnarray}
 $m\geq n$, $s+p\neq1$ and
the function $ g\left(y_1\cdots,\hat y_k,\cdots,\hat y_j,\cdots,y_n\right)$ does
       not depend on $y_k$ and $y_j$.
In our calculation, corollary 4 is used only in the case $s=0$ or $m$.

{\it proof}

We get 
\begin{eqnarray}
{}\makebox[-0.8cm]{}
&&\frac{ \left(y_j-1\right)^m y_j^p}{y_k-y_j}
g\left(y_1,\cdots,\hat y_k, \cdots,\hat y_j,\cdots,y_n\right)
\nonumber\\
{}\makebox[-2cm]{}
&\sim&
-\frac{ \left( q y_j-1\right)^m \left( q y_j\right)^p}{y_k- q y_j}
g\left(y_1,\cdots,\hat y_k, \cdots,\hat y_j,\cdots,y_n\right)
\nonumber\\
{}\makebox[-2cm]{}
&=&
\left(
-\frac{\left( y_k-1\right)^m  y_k^p}{y_k- q y_j}
+
\frac{\left( y_k-1\right)^m  y_k^p -\left( q y_j-1\right)^m \left( q y_j\right)^p}{y_k- q y_j}
\right)
g\left(y_1,\cdots,\hat y_k, \cdots,\hat y_j,\cdots,y_n\right)
\nonumber\\
{}\makebox[-2cm]{}
&\sim&
\left(
\frac{\left( q y_k-1\right)^m  \left( q y_k\right)^p}{ q\left(y_k-y_j\right)}
+
\frac{\left( y_k-1\right)^m  y_k^p -\left( q y_j-1\right)^m \left( q y_j\right)^p}{y_k- q y_j}
\right)
g\left(y_1,\cdots,\hat y_k, \cdots,\hat y_j,\cdots,y_n\right)
\nonumber\\
{}\makebox[-2cm]{}
&\sim&
\left(
\frac{\left( q y_j-1\right)^m  \left( q y_j\right)^p}{ q\left(y_k-y_j\right)}
+
\frac{\left( y_k-1\right)^m  y_k^p -\left( q y_j-1\right)^m \left( q y_j\right)^p}{y_k- q y_j}
\right)
g\left(y_1,\cdots,\hat y_k, \cdots,\hat y_j,\cdots,y_n\right).
\nonumber\\
{}\makebox[-2cm]{}
\end{eqnarray}
 The proposition 1 was used at the last weak equality, and the proposition
      2 was used at the other weak equalities.
Therefore, 
\begin{eqnarray}
&&
\frac{  \left(y_j-1\right)^m y_j^p-q^{p-1}\left( q y_j-1\right)^m  y_j^p}{ y_k-y_j}
g\left(y_1,\cdots,\hat y_k, \cdots,\hat y_j,\cdots,y_n\right)
\nonumber\\
&\sim&
\frac{\left( y_k-1\right)^m  y_k^p -\left( q y_j-1\right)^m \left( q y_j\right)^p}{y_k- q y_j}
g\left(y_1,\cdots,\hat y_k, \cdots,\hat y_j,\cdots,y_n\right)
\end{eqnarray}
holds.
Then,  expanding both numerators according to the formulae:
\begin{eqnarray}
  \left(y_j-1\right)^m y_j^p-q^{p-1}\left( q y_j-1\right)^m  y_j^p
&=&
\sum_{l=0}^{m}
b_ly_j^{l+p}
\\
\left( y_k-1\right)^m  y_k^p -\left( q y_j-1\right)^m \left( q y_j\right)^p
&=&
\left(y_k-q y_j\right)\sum_{l=0}^{m}
c_l
\sum_{r=0}^{l+r-1}
y_k^r\left(qy_j\right)^{l+p-r-1},
\label{eq:weak_equation_last}
\end{eqnarray}
we arrive at the formula (\ref{eq:corollary_2}).
\end{itemize}
The relations we wrote above are used to derive the ``canonical''
form of (\ref{eq:general_integrant}).

Next, we show two elementary integral formulas.
\begin{itemize}
\item The first integral formula is
\begin{eqnarray}
 \int_{C_-} \frac{e^{a x}}{\sinh^m x}dx
&=& 
-\frac{2 \pi i \prod_{j=1}^{m-1}\left(a-m+2j\right)}{\left(m-1\right)!
\left(e^{\left(a+m\right){\pi i}}-1\right)
}
\label{eq:value_of_integral_1}
\end{eqnarray}
 where $m$ is an integer and $-m<\Re(a)<m$.
\item The second integral formula is 
\begin{eqnarray}
&& \int_{C_-} \frac{e^{a x}}{
\sinh^m \left(x+y/2\right)\sinh^m\left(x-y/2\right)
}dx
\nonumber\\&=&
-\frac{2 \pi i 2^{2m-1}}{\left(m-1\right)!\left(e^{a \pi i}-1\right)}
\left(
\frac{\partial^{m-1}}{\partial X^{m-1}}
\left.\frac{X^{a/2+m-1}}{\left(X-e^{y}\right)^m}
\right|_{X=e^{-2y}}
+
\frac{\partial^{m-1}}{\partial X^{m-1}}
\left.\frac{X^{a/2+m-1}}{\left(X-e^{-y}\right)^m}
\right|_{X=e^{2y}}
\right),
\nonumber\\
\label{eq:value_of_integral_2}
\end{eqnarray}
where $m$ is an integer, $y$ is a real number and $-m<\Re(a)<m$.
Note that the l.h.s. of (\ref{eq:value_of_integral_2})
can be evaluated explicitly when $m$ is given explicitly.
For example, we have
\begin{eqnarray}
&& \int_{C_-} \frac{e^{a x}}{
\sinh^2 \left(x+y/2\right)\sinh^2 \left(x-y/2\right)
}dx
\nonumber\\&=&
\frac{ \pi i }{e^{a \pi i}-1}
\left(
-\frac{4 a         \cosh\left( {ya}/2\right)}{\sinh^2 y}
+\frac{8   \cosh y \sinh\left( {ya}/2\right)}{\sinh^3 y}
\right)
\label{eq:value_of_integral_2_1}
\end{eqnarray}
in the case $m=2$ and
\begin{eqnarray}
&& \int_{C_-} \frac{e^{a x}}{
\sinh^3 \left(x+y/2\right)\sinh^3 \left(x-y/2\right)
}dx
\nonumber\\&=&
\frac{ \pi i }{e^{a \pi i}-1}
\left(
-\frac{2\left(a^2+8\right)\sinh\left( {ya}/2\right)}{\sinh^3 y}
+\frac{12     a  \cosh  y \cosh\left( {ya}/2\right)}{\sinh^4 y}
-\frac{24                 \sinh\left( {ya}/2\right)}{\sinh^5 y}
\right)
\nonumber\\
\label{eq:value_of_integral_2_2}
\end{eqnarray}
in the case $m=3$,
etc..
\end{itemize} 
Now let us consider the integrals of a special form, namely, the integral 
\begin{eqnarray}
&& \prod_{j=1}^n
\int_{C_-}\frac{dx_j}{2\pi i}
U\left(x_1,\cdots,x_n\right)
P^{(n)}_{k} \left(e^{2x_1 \nu},e^{2x_3 \nu},\cdots,e^{2x_{2k-1} \nu}|
e^{2x_{2k+1} \nu},e^{2x_{2k+2} \nu},\cdots,e^{2x_{n} \nu}
\right) \nonumber \\
&& \hspace{5cm} \times \prod_{l=1}^{k} \frac{1}{e^{2x_{2l}}-e^{2x_{2l-1}}}, 
\end{eqnarray}
where ${P^{(n)}_{k}}$ is a polynomial of the form (\ref{eq:define_P_j}).
Using the relations
(\ref{eq:value_of_integral_1}) and (\ref{eq:value_of_integral_2}), we can 
integrate this expression $(n-k)$-times, namely with respect to
$(x_{2j-1}+x_{2j})/2$ and $x_{j'}$ where $j\leq k$,
$j'>2k$, while the variables $x_{2j-1}-x_{2j}$ are fixed. In this way, we can evaluate 
the $n$-dimensional integral  
\begin{eqnarray}
\prod_{j=1}^n
\int_{C_-}\frac{dx_j}{2\pi i}
U\left(x_1,\cdots,x_n\right)
T^c\left(\exp(2x_1 \nu),\cdots,\exp(2x_n \nu)\right),
\label{eq:canonical_integral_massless}
\end{eqnarray}
where $U$ is defined in (\ref{eq:definition_U}) and $T^c$ is a canonical
form function (\ref{eq:define_T^c}). It results into a polynomial with respect 
to one-dimensional integrals where the coefficients are rational functions
of $\sin \pi \nu$ and $\cos \pi \nu$.

\section{General discussion at the massive case}
The correlation functions  at massive case were obtained by Jimbo {\it et al} 
\cite{Jimbo92,Jimbo95}. Here we use the representation in \cite{Kitanine00},
\begin{eqnarray}
&& F\left[
\begin{array}{ccc}
\epsilon'_1 &\cdots &\epsilon'_n\\  
\epsilon _1 &\cdots &\epsilon _n
\end{array}
\right]
:=
 \left<
E^{\left(1\right)}_{\epsilon'_1 \epsilon_1}
\cdots
E^{\left(n\right)}_{\epsilon'_n \epsilon_n}
\right>\: \nonumber \\
&=&
\left(\phi i\right)^{-n(n-1)/2}
\prod_{j=1}^n
\int_{C'}\frac{dx_j}{2\pi i}
W\left(x_1,\cdots,x_n\right)
T\left(\exp(2x_1 \phi i),\cdots,\exp(2x_n \phi i)\right),
\label{eq:simplified_integral_form_P_n_massive}
\end{eqnarray}
where
$C'$ is the integral path from
${-\frac\pi2 (\phi^{-1}+i) }$ to ${\frac\pi2(\phi^{-1}-i) }$,
$T\left(y_1,\cdots,y_n\right)$ is defined at (\ref{eq:general_integrant}), 
\begin{eqnarray}
W\left( x_1,\cdots, x_n\right)
&=&
A_m
\vartheta_2( \prod_{m=1}^n z_m, q^{1/2})
\frac
{
\prod_{1\leq l<l'\leq n}
\vartheta_1(\frac{z_l}{ z_{l'}},q^{1/2})
}
{\prod_{l=1}^n\vartheta^n_1( z_l,q^{1/2})},
\nonumber\\
&&z_m\,\equiv\,\exp\left(\phi x_m i\right),
\quad  q\,=\,\exp\left(-\pi\phi\right),
\label{eq:definition_W}
\end{eqnarray}
\begin{eqnarray}
\vartheta_1\left( z,p\right)&\equiv&
-i p^{1/4}\left(z-z^{-1}\right)\prod_{m=1}^{\infty}
\left(1-p^{2m}z^2\right)
\left(1-p^{2m}z^{-2}\right)
\left(1-p^{2m}\right),
\\\vartheta_2\left( z,p\right)&\equiv&
p^{1/4}\left(z+z^{-1}\right)\prod_{m=1}^{\infty}
\left(1+p^{2m}z^2\right)
\left(1+p^{2m}z^{-2}\right)
\left(1-p^{2m}\right)
\end{eqnarray}
and
\begin{eqnarray}
 A_m&\equiv&
\phi^{n(n+1)/2}
\prod_{m=1}^\infty\left(\frac{1-q^m}{1+q^m}\right)^2\left[2 q^{1/8}
\prod_{m=1}^\infty\left(1-q^m\right)^3\right]^{\frac {m(m+1)}2-1}.
\end{eqnarray}
The parameter $\phi$ is defined by the relations $\Delta=\cosh \pi
\phi $ with $ 0\leq\phi $.

As in the massless case, we define weak equality
$\sim$.
Let us say that two functions $F(y_1,\cdots,y_n)$ and $G(y_1,\cdots,y_n)$ are 
weakly equivalent:
\begin{eqnarray}
 F\left(y_1,\cdots,y_n\right)&\sim&G\left(y_1,\cdots,y_n\right),
\end{eqnarray}
if
\begin{eqnarray}
\prod_{j=1}^n
\int_{C'}\frac{dx_j}{2\pi i}
W\left(x_1,\cdots,x_n\right)
\left[
 F\left(e^{2x_1 \nu},\cdots,e^{2x_n \nu}\right)
-G\left(e^{2x_1 \nu},\cdots,e^{2x_n \nu}\right)
\right]
&=&0.
\end{eqnarray}
Moreover, we use the term  ``canonical" in the same way as in the massless
case (\ref{eq:define_T^c}).

Below we show the correlation function $F\left[^{\epsilon'_1\,\cdots\,\epsilon'_n}_{\epsilon_1\,\cdots\,
 \epsilon_n}\right]$ for the massive case can be evaluated in the same way as  the massless case.
That is to say, we first obtain the canonical form, and then, reduce the multiple integral into 
the one-dimensional ones using this canonical form.

In fact, all the equations in the massless case,
i.e. (\ref{eq:transposition})$\sim$(\ref{eq:weak_equation_last}), also hold in
the massive case. The reason is that the function
$W\left(x_1,\cdots,x_n\right)$ has the following properties:
\begin{eqnarray}
W\left(x_1,\cdots,x_j,\cdots,x_k,\cdots,x_n\right)
&=&
-W\left(x_1,\cdots,x_k,\cdots,x_j,\cdots,x_n\right),\\
W\left(x_1,\cdots,x_j,\cdots,x_n\right)
&=&
-W\left(x_1,\cdots,  x_j+\pi i,\cdots,x_n\right).
\end{eqnarray}
$W\left(x_1,\cdots,x_n\right)$ is an elliptic function
 with periods $\pi i$ and $\pi/\phi$ with respect to each variable $x_j$.
$W\left(x_1,\cdots,x_n\right)$
 has  zero points at $x_j=x_k$ and a
 pole only at $x_j=0$, where $j\neq m$. 
 The order of the pole is $n$.
Therefore, the canonical forms for the correlation function in the massive 
case are the same as those for the massless case.

Next, we remark the relation
\begin{eqnarray}
 W\left(x_1,\cdots,x_n\right)
&=&
\sum_{m_1,\cdots,m_n=-\infty}^{\infty}U\left(x_1+\frac{\pi m_1}\phi  ,\cdots,x_n+\frac{\pi m_n}\phi  \right)
\label{eq:relation_W_U}
\end{eqnarray}
 holds. The proof is as follows.
The both sides of (\ref{eq:relation_W_U}) have following properties.
 First, they are elliptic functions with periods $\pi/\phi$ and $2 \pi
 i$ with respect to each variable $x_j$.
Second, the functions have poles only at
 $x_j=k\pi/\phi+ l \pi i$ 
and zeros at
  $x_j=k\pi/\phi+ l \pi i+ x_m$, where $k$ and $l$ are arbitrary integers and $j\neq m$. 
Third,
the order of the pole is $n$ and the order of any zero point is $1$.
Finally, both values 
\begin{eqnarray}
\lim_{x_1\rightarrow0}
\lim_{x_2\rightarrow0}
\cdots
\lim_{x_n\rightarrow0}
 W\left(x_1,\cdots,x_n\right)
\prod_{j=1}^{n}{x_j}^j
\end{eqnarray}
and
\begin{eqnarray}
\lim_{x_1\rightarrow0}
\lim_{x_2\rightarrow0}
\cdots
\lim_{x_n\rightarrow0}
 \sum_{m_1,\cdots,m_n=-\infty}^{\infty}U\left(x_1+\frac{\pi m_1}\phi  ,\cdots,x_n+\frac{\pi m_n}\phi  \right)
\prod_{j=1}^{n}{x_j}^j
 \end{eqnarray}
 are 1.

By the relation (\ref{eq:relation_W_U}), we find
\begin{eqnarray}
&&\prod_{j=1}^n
\int_{C'}\frac{dx_j}{2\pi i}
W\left(x_1,\cdots,x_n\right)
f\left(\exp\left(2 x_1\phi i\right),\cdots,\exp\left(2 x_n\phi i\right)\right)
\nonumber\\&=&
\prod_{j=1}^n
\int_{C_j}\frac{dx_j}{2\pi i}
U\left(x_1,\cdots,x_n\right)
f\left(\exp\left(2 x_1\phi i\right),\cdots,\exp\left(2 x_n\phi i\right)\right)
\label{eq:relation_W_U_integral_form}
\end{eqnarray}
holds, where $C_j$ is the integral path from $-\infty-(1/2-j\delta)\pi
i$ to $\infty-(1/2-j\delta)\pi i$ and the both integrands have no pole on
the integral paths respectively.
Then, using the relation 
(\ref{eq:value_of_integral_1}) and (\ref{eq:value_of_integral_2}), 
one can conclude, 
 the multiple-integral
\begin{eqnarray}
\prod_{j=1}^n
\int_{C'}\frac{dx_j}{2\pi i}
W\left(x_1,\cdots,x_n\right)
T^c\left(\exp(i 2x_1 \phi),\cdots,\exp(i 2x_n \phi)\right),
\label{eq:canonical_integral_massive}
\end{eqnarray}
where $W$ is defined as (\ref{eq:definition_W}) and $T^c$ is a canonical form, 
reduces to a polynomial with respect to one-dimensional integral. Note that 
here the coefficients of the polynomial are rational functions of
$\cosh\pi \phi$ and $\sinh\pi \phi$.

\section{In the case of $F\left[^{++}_{++}\right]$}
We illustrate how the procedures in the previous sections work  for the simplest case 
$F\left[^{++}_{++}\right]$, i.e.,
\begin{eqnarray}
\nu^{-1}
\int_{C_-}\frac{dx_1}{2\pi i}
\int_{C_-}\frac{dx_2}{2\pi i}
U\left(x_1,x_2\right)
T\left(\exp(2x_1 \nu),\exp(2x_2 \nu)\right),
\end{eqnarray}
in the case  $-1<\Delta\leq 1$, and 
\begin{eqnarray}
-i\phi^{-1}
\int_{C'}\frac{dx_1}{2\pi i}
\int_{C'}\frac{dx_2}{2\pi i}
W\left(x_1,x_2\right)
T\left(\exp(2x_1 \nu),\exp(2x_2 \nu)\right),
\end{eqnarray}
in the case $1< \Delta$. Here $U\left(x_1,x_2\right)$ and $W\left(x_1,x_2\right)$ are defined in
(\ref{eq:definition_U}) and 
(\ref{eq:definition_W}) respectively.
For $F\left[^{++}_{++}\right]$, it is very simple to perform the first step, 
namely, to get the ``canonical'' form (\ref{eq:define_T^c})
 because we do not need to reduce a power of
denominator. Indeed,
\begin{eqnarray}
 T(y_1,y_2)&=&
\frac
{
\left(y_1-1\right)
\left(qy_2 -1\right)
}
{2\left(y_1-y_2 q\right)}
\;=\;
\frac{ q y_2- 1}2+\frac
{
\left(qy_2 -1\right)^2
}
{2\left(y_1-qy_2 \right)}
\;\sim \;
\frac q 2 y_2
-\frac{\left(y_2-1\right)^2}
{2\left(y_1-y_2\right)},
\label{eq:T_tmp_1}
\end{eqnarray}
where we have used the relations (\ref{eq:transposition}) and 
(\ref{eq:proposition1}). Then, using the relation (\ref{eq:corollary_2}) 
for $m=2$, we have
\begin{eqnarray}
-\frac{
\left(y_2-1\right)^2}{2\left(y_1-y_2\right)}
&=&-\frac{y_2^2}{2\left(y_1-y_2\right)}
+\frac{ y_2}{y_1-y_2}
-\frac{1}{2\left(y_1-y_2\right)}
\nonumber\\
&\sim&
-\frac{1}{2 q\left(y_1-y_2\right)}
+\frac{y_2}2
+\frac{ y_2}{y_1-y_2}
-\frac{1}{2\left(y_1-y_2\right)}
\nonumber\\&=&
\frac{2 y_2-1- q^{-1}}{2\left(y_1-y_2\right)}
+\frac{y_2}2.
\end{eqnarray}
Substituting it to the r.h.s. of (\ref{eq:T_tmp_1}), we get
\begin{eqnarray}
T(y_1,y_2)&\sim&T^c\left(y_1,y_2\right)
\;=\;
\frac{ q +1}2 y_2+\left(y_2-\frac{ q+1}{2 q}\right)\frac1{y_1-y_2}
\label{eq:canonical_form_P2}
\end{eqnarray}
and this is the ``canonical'' form for $T$, i.e. the polynomials
$P^{(2)}_0$ and $P^{(2)}_1$ at (\ref{eq:define_T^c}) are equal to $( q+1)y_2/2$ and $y_2-(1+ q^{-1})/2$
respectively.

Let us consider an integral:
\begin{eqnarray}
&&
\nu^{-1}
\int_{C_-}\frac{dx_1}{2\pi i}
\int_{C_-}\frac{dx_2}{2\pi i}
\frac
{
\sinh\left(x_1-x_2\right)
}
{
\sinh^2 x_1 \sinh^2 x_2
}
\frac{ q+1}{2}e^{2x_2\nu}
\nonumber\\
&=&
\frac{ q+1}{4\left(2\pi i\right)^2\nu}
\int_{C_-}dx_1
\int_{C_-}dx_2
\frac
{
e^{x_1}e^{(2\nu-1)x_2}
-e^{-x_1}e^{(2\nu+1)x_2}
}
{
\sinh^2 x_1 \sinh^2 x_2
}
\nonumber\\
&=&
\frac{ q+1}{4\left(2\pi i\right)^2\nu}
\left[
\left( \pi i \right)
\left(\frac{2\pi i \left(2\nu-1\right)}{ q+1}\right)
-
\left(-\pi i\right)
\left(\frac{2\pi i \left(2\nu+1\right)}{ q+1}\right)
\right]
\nonumber\\
&=&\frac12.
\label{eq:example_integral_1_massless}
\end{eqnarray}
This integrand corresponds to $U\left(x_1,x_2\right)$ times the first term of 
(\ref{eq:canonical_form_P2}). Here we used the formula (\ref{eq:value_of_integral_1}) at the second equality.
Next, we evaluate another integral:
\begin{eqnarray}
&&
\nu^{-1}
\int_{C_-}\frac{dx_1}{2\pi i}
\int_{C_-}\frac{dx_2}{2\pi i}
\frac
{
\sinh\left(x_1-x_2\right)
}
{
\sinh^2 x_1 \sinh^2 x_2
}
\left(e^{2x_2\nu}-\frac{ q+1}{2 q}\right)\frac1{e^{2x_1\nu}-e^{2x_2\nu}}
\nonumber\\
&=&
\frac{1}{2\left(2 \pi i\right)^2\nu}
\int_{-\infty}^{\infty}dy
\frac{\sinh y}{\sinh \nu y}
\int_{C_-}dx
\frac
{
e^{-\nu y}-\frac{q+1}{2q}e^{-2\nu x}
}
{
\sinh^2 \left(x+y/2\right) \sinh^2 \left(x-y/2\right)
}
\nonumber\\
&=&
\frac1{2\pi^2\nu}\int_{-\infty}^{\infty}dy
\frac{1}{\sinh y}
\left(
\frac{e^{\nu y}}{ \sinh\nu y}
-
\frac{ i\pi\nu\left(q+1\right)\cosh\nu y}{\left(q-1\right) \sinh\nu y}
\right)
+
\frac{\cosh y}{\sinh^2 y}
\left(
\frac{ i\pi\left(q+1\right)}{\left(q-1\right)}
-
\frac{ y e^{\nu y} }{\sinh\nu y} 
\right)
\nonumber\\
&=&
\int_{C_-}dy
\frac{1}{\sinh y}
\left(
-
\frac{\cos \pi\nu}{2\pi\sin \pi\nu}
\frac{\cosh\nu y}{ \sinh\nu y}
-
\frac{ 1  }{2 \pi^2 }
\frac\partial{\partial \nu}\frac{\cosh\nu y}{ \sinh\nu y}
\right).
\label{eq:example_integral_2_massless}
\end{eqnarray}
This integrand corresponds to $U\left(x_1,x_2\right)$ times the second term of (\ref{eq:canonical_form_P2}).
At the second equality, we used the formula
(\ref{eq:value_of_integral_2_1}), and at the third equality we used the
relation
\begin{eqnarray}
 \int_{C_-} dy\frac{\cosh y}{\sinh^2 y}f\left(y\right)
&=& \int_{C_-} dy\frac{1}{\sinh y}\frac {d}{dy}f\left(y\right),
\label{eq:simple_relation}
\end{eqnarray}
where $\lim_{y\rightarrow \pm \infty-i\pi/2} f(y)/\sinh y=0$.

As for the massive case,  by exchanging $U\left(x_1,x_2\right)$ and $C_-$ with $W\left(x_1,x_2\right)$ 
and $C'$, we can calculate as
\begin{eqnarray}
-i\phi^{-1}
\int_{C'}\frac{dx_1}{2\pi i}
\int_{C'}\frac{dx_2}{2\pi i}
W\left(x_1,x_2\right)
\frac{ q+1}{2}e^{2x_2\phi i}
&=&\frac12,
\label{eq:example_integral_1_massive}
\end{eqnarray}
and
\begin{eqnarray}
&&-i\phi^{-1}
\int_{C'}\frac{dx_1}{2\pi i}
\int_{C'}\frac{dx_2}{2\pi i}
W\left(x_1,x_2\right)
\left(e^{2x_2\phi i}-\frac{ q+1}{2 q}\right)\frac1{e^{2x_1\phi i}-e^{2x_2\phi i}}
\nonumber\\
&=&
\int_{C_-}dy
\frac{1}{\sinh y}
\left(
-
\frac{\cos \pi\phi i}{2\pi\sin \pi\phi i}
\frac{\cosh\phi i y}{ \sinh\phi i y}
-
\frac{ 1  }{2 \pi^2 }
\frac\partial{\partial \phi i}\frac{\cosh\phi i y}{ \sinh\phi i y}
\right).
\label{eq:example_integral_2_massive}
\end{eqnarray}
Actually, once we used the relation (\ref{eq:relation_W_U_integral_form}), 
we see the rest of the derivation is the same as the massless case. 
Therefore, the final results of these two cases are very similar. 
More precisely speaking, the results can be exchanged to each other 
with the exchange of $\phi i$ and $\nu$.
This property always holds in the evaluation of the other correlation functions.

Gathering all the results, we get the following expression
\begin{eqnarray}
 F\left[^{++}_{++}\right]&=&\frac12
-\frac{\cos \pi\eta}{2\pi\sin \pi\eta}
\int_{C_-}dy
\frac{1}{\sinh y}
\frac{\cosh\eta y}{ \sinh\eta y}
-
\frac{ 1  }{2 \pi^2 }
\int_{C_-}dy
\frac{1}{\sinh y}
\frac\partial{\partial \eta}\frac{\cosh\eta y}{ \sinh\eta y},
\end{eqnarray}
where $\eta$ is equal to $\nu$ in the case $-1<\Delta=\cos\pi\nu\leq1$
and is equal to $\phi i$ in the case $1\leq\Delta=\cos\pi\phi$.

\section{The third-neighbor correlation functions and some other
 correlation functions}
Using the method in the previous sections, we have analyzed the
all correlation functions 
$F[^{\epsilon'_1,\cdots,\epsilon'_n}_{\epsilon_1,\cdots,\epsilon_n}]$
for $n\leq4$. Actually the derivations of the canonical form from the multiple 
integral representations of the correlation functions are very similar to each 
other. So, in \ref{sec:example}, we explain the outline of its derivation for 
$F[^{++++}_{++++}]$ as an example. Also the rest of our tasks, i.e. the evaluation 
of the integrals with respect to the canonical forms, are straightforward 
if one uses the relations (\ref{eq:value_of_integral_1}), (\ref{eq:value_of_integral_2})
and the integration by parts like (\ref{eq:simple_relation}). Then, we omit the account 
of these calculations and give the final result in \ref{sec:results} together with other 
independent results up to ${n=4}$. 
The correlation functions which are not given in Appendix B explicitly 
should be derived from the relations, 
\begin{eqnarray}
& F[^{ \epsilon'_1, \epsilon'_ 2   ,\cdots, \epsilon'_n  }
    _{ \epsilon _1, \epsilon _ 2   ,\cdots, \epsilon _n  }]
\;=\;
  F[^{ \epsilon'_1, \epsilon'_ 2   ,\cdots, \epsilon'_n,+}
    _{ \epsilon _1, \epsilon _ 2   ,\cdots, \epsilon _n,+}]
+ F[^{ \epsilon'_1, \epsilon'_ 2   ,\cdots, \epsilon'_n,-}
    _{ \epsilon _1, \epsilon _ 2   ,\cdots, \epsilon _n,-}],
\\
& F[^{ \epsilon'_1, \epsilon'_ 2   ,\cdots, \epsilon'_n  }
    _{ \epsilon _1, \epsilon _ 2   ,\cdots, \epsilon _n  }]
\;=\;
  F[^{ \epsilon'_n, \epsilon'_{n-1},\cdots, \epsilon'_1  }
    _{ \epsilon _n, \epsilon _{n-1},\cdots, \epsilon _1  }]
\;=\;
  F[^{ \epsilon _1, \epsilon _ 2   ,\cdots, \epsilon _n  }
    _{ \epsilon'_1, \epsilon'_ 2   ,\cdots, \epsilon'_n  }]
\;=\;
  F[^{-\epsilon'_1,-\epsilon'_ 2   ,\cdots,-\epsilon'_n  }
    _{-\epsilon _1,-\epsilon _ 2   ,\cdots,-\epsilon _n  }],
\\
& F[^{ \epsilon'_1, \epsilon'_ 2   ,\cdots, \epsilon'_n  }
    _{ \epsilon _1, \epsilon _ 2   ,\cdots, \epsilon _n  }]\;=\;0\quad\quad\makebox{in case $\#\{j|\epsilon'_j=+\}\neq\#\{j|\epsilon_j=+\}$}.
\end{eqnarray}

\begin{figure}
\begin{center}
\begin{psfrags}
      \psfrag{a}{\tiny $\left<S_j^x S_{j+3}^x\right>$}
      \psfrag{b}{\tiny $\left<S_j^z S_{j+3}^z\right>$}
      \psfrag{z}{ $\Delta$}
  \psfig{file=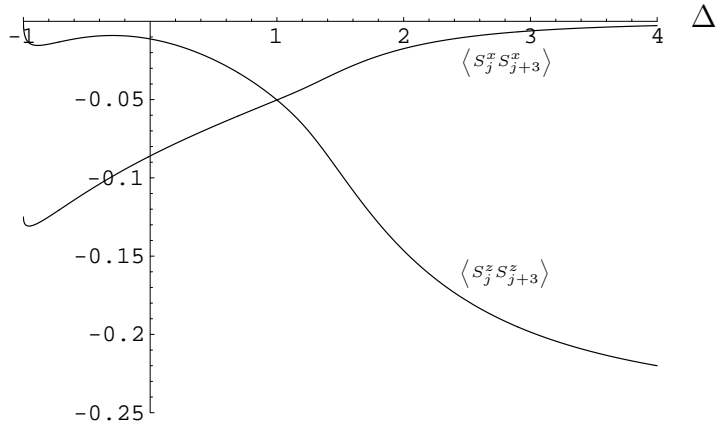 , scale = 0.9}
\caption{The third-neighbor correlation functions for the $XXZ$ chain.}
\label{graph:first}
\end{psfrags}
\end{center}
\end{figure}

From our results, we can express many correlation functions as a 
polynomial with respect to one-dimensional integrals. For example, 
the third-neighbor correlation functions
$\left<S_j^xS_{j+3}^x\right>$ and 
 $\left<S_j^zS_{j+3}^z\right>$ are given by 
\begin{eqnarray}
&&\left<S_j^xS_{j+3}^x\right> 
\nonumber\\
&=&
 \frac{3 }{4 \pi  \left( 1 + 2 {c_2} \right)  {s_1}}
       {\zeta _\eta}(1)
+\frac{{c_1} \left( -1 + 2 {c_2} \right) }{4 {\pi }^2}
       {\zeta'_\eta}(1)
-\frac{\left( -1 + 7 {c_2} + 4 {c_4} \right)  {s_1}}{2 \pi  \left( 1 + 2 {c_2} \right) }
       {\zeta _\eta}(3)
-\frac{{c_1} {{s_1}}^2 }{2 {\pi }^2}
       {\zeta'_\eta}(3)
\nonumber\\&&
-\frac{5 \left( 2 {c_2} + {c_4} \right)  {s_1}}{4 \pi  \left( 1 + 2 {c_2} \right) }
       {\zeta _\eta}(5)
-\frac{{c_1} {c_2} {{s_1}}^2 }{4 {\pi }^2}
       {\zeta'_\eta}(5)
-\frac{{c_1} \left( -12 + 4 {c_2} - 7 {c_4} \right)}{4 {\pi }^2 \left( 1 + 2 {c_2} \right) }
       {\zeta _\eta}(1) {\zeta _\eta}(3)
\nonumber\\&&
+\frac{\left( 6 + 3 {c_2} + {c_4} \right)  {s_1}}{8 {\pi }^3}
       {\zeta'_\eta}(1) {\zeta _\eta}(3)
+\frac{\left( -1 + 3 {c_2} \right)  {s_1}}{8 {\pi }^3}
       {\zeta _\eta}(1) {\zeta'_\eta}(3)
+\frac{{{c_1}}^3 {{s_1}}^2 }{4 {\pi }^4}
       {\zeta'_\eta}(1) {\zeta'_\eta}(3)
\nonumber\\&&
+\frac{5 {c_1} \left( 9 + 4 {c_2} + 5 {c_4} \right) }{8 {\pi }^2 \left( 1 + 2 {c_2} \right) }
       {\zeta _\eta}(1) {\zeta _\eta}(5)
+\frac{5 {{c_1}}^2 \left( 2 + {c_2} \right)  {s_1}}{4 {\pi }^3}
       {\zeta'_\eta}(1) {\zeta _\eta}(5)
\nonumber\\&&
+\frac{{{c_1}}^2 \left( 2 + {c_2} \right)  {s_1}}{4 {\pi }^3}
       {\zeta _\eta}(1) {\zeta'_\eta}(5)
+\frac{3 {{c_1}}^3 {{s_1}}^2}{4 {\pi }^4}
       {\zeta'_\eta}(1) {\zeta'_\eta}(5)
-\frac{3 {c_1} \left( 9 + 4 {c_2} + 5 {c_4} \right) }{16 {\pi }^2 \left( 1 + 2 {c_2} \right) }
      {{\zeta _\eta}(3)}^2
\nonumber\\&&
-\frac{{{c_1}}^2 \left( 2 + {c_2} \right)  {s_1}}{4 {\pi }^3}
       {\zeta _\eta}(3) {\zeta'_\eta}(3)
-\frac{{{c_1}}^3 {{s_1}}^2}{8 {\pi }^4} 
      {{\zeta'_\eta}(3)}^2 
\label{eq:transverse_third_neighbor}
\\
&&\left<S_j^zS_{j+3}^z\right>
\nonumber\\&=&
\frac{1}{4}
-\frac{3 {c_1} \left( -1 + 2 {c_2} \right)}{2 \pi \left( 1 + 2 {c_2} \right) {s_1}}
   {\zeta _\eta}(1)
-\frac{1}{2 {\pi }^2}
   {\zeta'_\eta}(1)
+\frac{\left( 3 + 2 {c_2} \right) \left( 7 + {c_4} \right) {s_1}}{4 \pi {c_1} \left( 1 + 2 {c_2} \right) }
   {\zeta _\eta}(3)
+\frac{{{c_1}}^2 {{s_1}}^2 }{{\pi }^2}
   {\zeta'_\eta}(3)
\nonumber\\&&
+\frac{5 \left( 5 + 3 {c_2} + 3 {c_4} + {c_6} \right) {s_1} }{8 \pi {c_1} \left( 1 + 2 {c_2} \right) }
   {\zeta _\eta}(5)
+\frac{{c_2} {{s_1}}^2 }{2 {\pi }^2}
   {\zeta'_\eta}(5) 
-\frac{\left( 2 + 23 {c_2} + 4 {c_4} + {c_6} \right)}{4 {\pi }^2 \left( 1 + 2 {c_2} \right) }
   {\zeta _\eta}(1) {\zeta _\eta}(3)
\nonumber\\&&
-\frac{{c_1} \left( 4 + {c_2} \right) {s_1} }{2 {\pi }^3}
   {\zeta'_\eta}(1) {\zeta _\eta}(3)
-\frac{{c_1} {c_2} {s_1} }{2 {\pi }^3}
   {\zeta _\eta}(1) {\zeta'_\eta}(3)
-\frac{{{c_1}}^2 {{s_1}}^2}{2 {\pi }^4}
   {\zeta'_\eta}(1) {\zeta'_\eta}(3)
\nonumber\\&&
-\frac{5 \left( 8 + 23 {c_2} + 4 {c_4} + {c_6} \right) }{8 {\pi }^2 \left( 1 + 2 {c_2} \right) }
   {\zeta _\eta}(1) {\zeta _\eta}(5)
-\frac{5 {c_1} \left( 2 + {c_2} \right) {s_1} }{2 {\pi }^3}
   {\zeta'_\eta}(1) {\zeta _\eta}(5)
\nonumber\\&&
-\frac{{c_1} \left( 2 + {c_2} \right) {s_1}}{2 {\pi }^3}
   {\zeta _\eta}(1) {\zeta'_\eta}(5)
-\frac{3 {{c_1}}^2 {{s_1}}^2 }{2 {\pi }^4}
   {\zeta'_\eta}(1) {\zeta'_\eta}(5)
+\frac{3 \left( 8 + 23 {c_2} + 4 {c_4} + {c_6} \right) }{16 {\pi }^2 \left( 1 + 2 {c_2} \right) }
  {{\zeta _\eta}(3)}^2
\nonumber\\&&
+\frac{{c_1} \left( 2 + {c_2} \right) {s_1} }{2 {\pi }^3}
   {\zeta _\eta}(3) {\zeta'_\eta}(3)
+\frac{{{c_1}}^2 {{s_1}}^2 }{4 {\pi }^4}
  {{\zeta'_\eta}(3)}^2.
\label{eq:longitudinal_third_neighbor}
\end{eqnarray}
Here we used following notations to make the results short : 
\begin{eqnarray}
 c_j&:=&\cos \pi j\eta,
\label{ea:notation_first}\\
 s_j&:=&\sin \pi j\eta, \\
\zeta _\eta\left(j\right)&:=&\int_{C_-}dx\frac1{\sinh x}\frac{\cosh \eta x}{ \sinh^j \eta x},
\label{ea:notation_integral}\\
\zeta'_\eta\left(j\right)&:=&\int_{C_-}dx\frac1{\sinh x}\frac{\partial}{\partial\eta}\frac{\cosh \eta x}{ \sinh^j \eta x},
\label{ea:notation_last}
\end{eqnarray}
where $C_-$ is an integral path from $-\infty-\pi i/2$ to  $\infty-\pi i/2$
and $\eta$ is equal to $\nu$ in case $-1<\Delta=\cos\pi\nu\leq1$
and is equal to $\phi i$ in case $1<\Delta=\cos\pi\phi$. Note that we have used the 
relations
\begin{eqnarray}
\left<S_j^xS_{j+3}^x\right>&=& 
 F\left[^{+++-}_{-+++}\right]+ F\left[^{++--}_{-+-+}\right],
\\
\left<S_j^zS_{j+3}^z\right>&=&
 -2 F\left[^{+-+-}_{+-+-}\right]
 +2 F\left[^{++++}_{++++}\right]
 -3 F\left[^{++}_{++}\right]+\frac34, 
\end{eqnarray}
to get the results (\ref{eq:transverse_third_neighbor}) and (\ref{eq:longitudinal_third_neighbor}).
The expressions (\ref{eq:transverse_third_neighbor}) and (\ref{eq:longitudinal_third_neighbor}) allow 
us to give the numerical values of the correlation functions with a very high precision. They are plotted 
in Fig. \ref{graph:first}. Furthermore, when $\eta$ is a real rational number, the one-dimensional integrals 
(\ref{ea:notation_integral}) and (\ref{ea:notation_last}) can be integrated analytically and we can get 
the complete analytical values for (\ref{eq:transverse_third_neighbor}) and (\ref{eq:longitudinal_third_neighbor}). 
Some explicit examples are given in Table \ref{table:exact_values}.

\begin{figure}
\begin{center}
\begin{psfrags}
      \psfrag{a}{\tiny $\left<S_j^zS_{j+1}^zS_{j+2}^z S_{j+3}^z\right>$}
      \psfrag{b}{\tiny $\left<S_j^xS_{j+1}^xS_{j+2}^z S_{j+3}^z\right>$}
      \psfrag{c}{\tiny $\left<S_j^xS_{j+1}^xS_{j+2}^x S_{j+3}^x\right>$}
      \psfrag{d}{\tiny $\left<S_j^xS_{j+1}^xS_{j+2}^y S_{j+3}^y\right>$}
      \psfrag{z}{ $\Delta$}
  \psfig{file=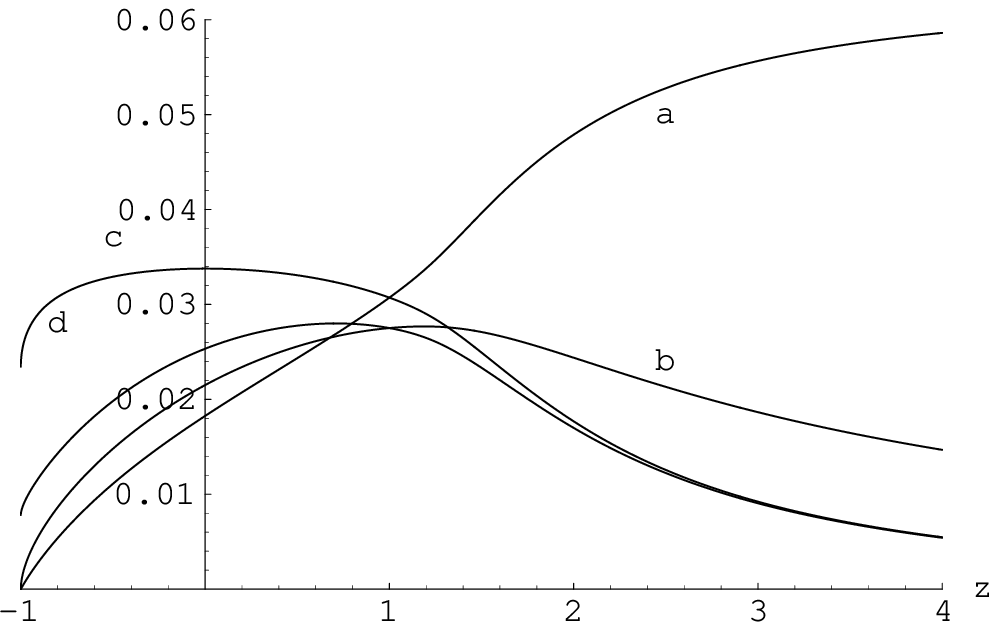 , scale = 0.9}
\caption{Four-point correlation functions,
  $\left<S_j^zS_{j+1}^zS_{j+2}^z S_{j+3}^z\right>$
  $\left<S_j^xS_{j+1}^xS_{j+2}^z S_{j+3}^z\right>$
  $\left<S_j^xS_{j+1}^xS_{j+2}^x S_{j+3}^x\right>$
   $\left<S_j^xS_{j+1}^xS_{j+2}^y S_{j+3}^y\right>$
  , for the $XXZ$ chain.}
\label{graph:second}
\end{psfrags}
\begin{psfrags}
      \psfrag{a}{\tiny $\left<S_j^zS_{j+1}^xS_{j+2}^x S_{j+3}^z\right>$}
      \psfrag{b}{\tiny $\left<S_j^xS_{j+1}^yS_{j+2}^y S_{j+3}^x\right>$}
      \psfrag{c}{\tiny $\left<S_j^xS_{j+1}^zS_{j+2}^z S_{j+3}^x\right>$}
      \psfrag{d}{\tiny $\left<S_j^xS_{j+1}^zS_{j+2}^x S_{j+3}^z\right>$}
      \psfrag{e}{\tiny $\left<S_j^xS_{j+1}^yS_{j+2}^x S_{j+3}^y\right>$}
      \psfrag{z}{ $\Delta$}
  \psfig{file=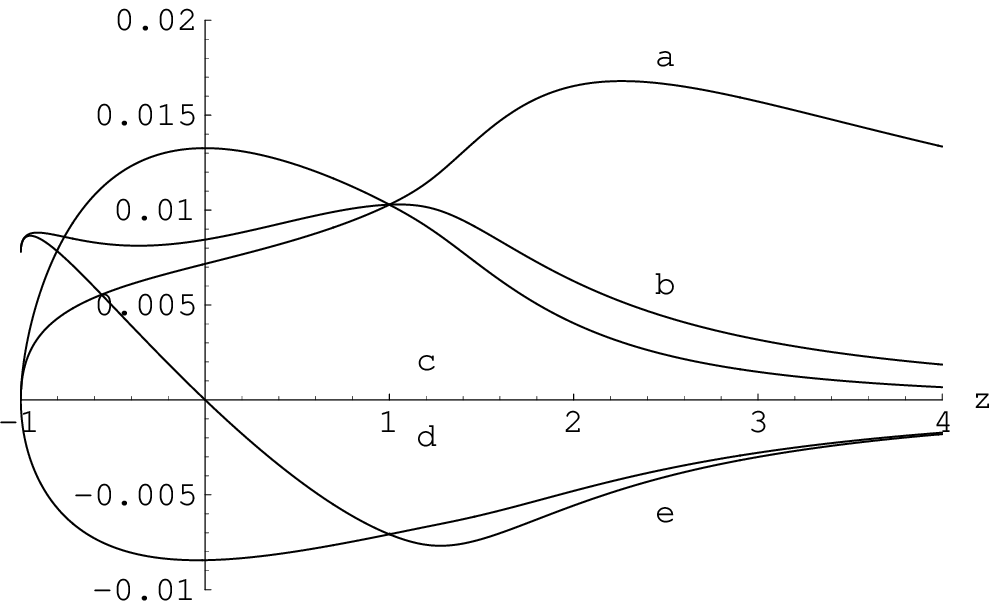 , scale = 0.9}
\end{psfrags}
\caption{Four-point correlation functions,
 $\left<S_j^zS_{j+1}^xS_{j+2}^x S_{j+3}^z\right>$
 $\left<S_j^xS_{j+1}^yS_{j+2}^y S_{j+3}^x\right>$
 $\left<S_j^xS_{j+1}^zS_{j+2}^z S_{j+3}^x\right>$
 $\left<S_j^xS_{j+1}^zS_{j+2}^x S_{j+3}^z\right>$
 $\left<S_j^xS_{j+1}^yS_{j+2}^x S_{j+3}^y\right>$
  , for the $XXZ$ chain.}
\label{graph:third}
\end{center}
\end{figure}

\begin{table}
 \begin{tabular}{clll}
\hline
\hline
\\[-0.4cm]
 $\nu$ 
& $\left<S_j^zS_{j+2}^z\right>$
&$\left<S_j^zS_{j+3}^z\right>$
&$\left<S_j^xS_{j+3}^x\right>$
\\[-0.4cm]
\\\hline

\\[-0.4cm]$ 0$

&$\frac{1}{12} - \frac{4 \log (2)}{3} + \frac{3 \zeta(3)}{4}$
&$\frac{1}{12} - 3 \log (2) + \frac{37 \zeta(3)}{6} - \frac{125 \zeta(5)}{24}-$
&$\frac{1}{12} - 3 \log (2) + \frac{37 \zeta(3)}{6} - \frac{125 \zeta(5)}{24}-$
\\
&
&$\frac{14 \log (2) \zeta(3)}{3} + \frac{25 \log (2) \zeta(5)}{3} - \frac{3 {\zeta(3)}^2}{2}$
&$\frac{14 \log (2) \zeta(3)}{3} + \frac{25 \log (2) \zeta(5)}{3} - \frac{3 {\zeta(3)}^2}{2}$

\\\\[-0.4cm]$ \frac{1}{2}$

&$0$
&$-\frac{1}{9 {\pi }^2}$
&$-\frac{8}{3 {\pi }^3}$

\\\\[-0.4cm]$ \frac{1}{3}$

&$\frac{7}{256} $
&$- \frac{ 401}{16384}$
&$- \frac{4399}{65536}$

\\\\[-0.4cm]$ \frac{2}{3}$

&$\frac{8671}{4096} - \frac{49 {\sqrt{3}}}{64} - \frac{1305}{512 \pi } $
&$ - \frac{25162841}{ 4194304}  + \frac{703383 {\sqrt{3}}}{131072} + \frac{1018791}{ 262144 \pi }-$
&$ - \frac{23685209}{16777216}  + \frac{113127 {\sqrt{3}}}{131072} + \frac{2138535}{1048576 \pi }-$
\\&
&$\frac{7533 {\sqrt{3}}}{1024 \pi } - \frac{19683}{ 4096 {\pi }^2}$
&$\frac{2673 {\sqrt{3}}}{2048 \pi } - \frac{19683}{16384 {\pi }^2}$

\\\\[-0.4cm]$ \frac{1}{4}$

&$- \frac{5}{8} + \frac{4}{\pi } - \frac{6}{{\pi }^2}$ 
&$- \frac{39}{16} + \frac{22}{\pi } - \frac{677}{9 {\pi }^2} + \frac{1088}{9 {\pi }^3} - \frac{256}{3 {\pi }^4}$
&$\frac{35{\sqrt{2}}}{64 } - \frac{131 {\sqrt{2}}}{24 \pi } + \frac{707 {\sqrt{2}}}{36 {\pi }^2} - \frac{296 {\sqrt{2}}}{9 {\pi }^3} +$
\\&&&$ \frac{64 {\sqrt{2}}}{3 {\pi }^4}$

\\\\[-0.4cm]$ \frac{3}{4}$

&$\frac{36169}{17496} - \frac{160{\sqrt{3}}}{243 } - \frac{4}{3 \pi }-$
&$ - \frac{2970584653}{229582512} + \frac{41685488 {\sqrt{3}}}{ 4782969 }-$
&$ - \frac{3334212769  {\sqrt{2}}}{ 918330048} + \frac{11666693 {\sqrt{6}}}{4782969} +$
\\
&$ \frac{608{\sqrt{3}}}{729  \pi } - \frac{2}{3 {\pi }^2}$
&$\frac{22418}{531441 \pi } + \frac{3638960 {\sqrt{3}}}{4782969 \pi }-$
&$ \frac{235397 {\sqrt{2}}}{ 4251528 \pi } - \frac{95308 {\sqrt{6}}}{ 4782969 \pi } -$
\\&
&$\frac{2541253}{531441 {\pi }^2} - \frac{197120 {\sqrt{3}}}{19683 {\pi }^2}-$
&$ \frac{2738083{\sqrt{2}}}{ 2125764 {\pi }^2} - \frac{52544 {\sqrt{6}}}{  19683 {\pi }^2} -$
\\&
&$\frac{1088}{243 {\pi }^3} - \frac{188416 {\sqrt{3}}}{ 59049  {\pi }^3} - \frac{256}{243 {\pi }^4}$
&$ \frac{296 {\sqrt{2}}}{243 {\pi }^3} - \frac{47104 {\sqrt{6}}}{ 59049{\pi }^3} - \frac{64 {\sqrt{2}}}{ 243{\pi }^4}$

\\\\[-0.4cm]$ \frac{1}{5}$

&$- \frac{3529}{512} + \frac{1589 {\sqrt{5}}}{512}$
&$- \frac{7128183}{32768}   + \frac{3187253 {\sqrt{5}}}{32768}$
&$\frac{6215287}{131072} - \frac{2782879 {\sqrt{5}}}{131072}$

\\\\[-0.4cm]$ \frac{1}{6}$

&$- \frac{283}{288} + \frac{11{\sqrt{3}}}{3\pi }- \frac{39}{4 {\pi }^2}$
&$- \frac{622483}{62208}  +  \frac{5216{\sqrt{3}}}{81  \pi } - \frac{69007}{144 {\pi }^2} +$
&$ \frac{202393 {\sqrt{3}}}{82944} -  \frac{27697}{576 \pi } + \frac{23045 {\sqrt{3}}}{ 192  {\pi }^2}$
\\&
&$ \frac{546 {\sqrt{3}}}{{\pi }^3} - \frac{729}{{\pi }^4}$
&$ - \frac{1647}{4 {\pi }^3} + \frac{729 {\sqrt{3}}}{4 {\pi }^4}$

\\\\[-0.4cm]$1$

&$0$
&$0$
&$- \frac{1}{8}$
\\
\\[-0.4cm]
\hline
\hline
 \end{tabular}
\caption{Exact values of the third-neighbor correlation functions at
 several points of ${\Delta}$. For comparison, we also give the values for the second-neighbor correlations here. }
\label{table:exact_values}
\end{table}
 The polynomial representations of other correlation functions, $\left<S_j^zS_{j+1}^zS_{j+2}^z
S_{j+3}^z\right>$ etc., can be derived similarly. Here we simply show 
the shapes of the correlation functions in Fig. \ref{graph:second} and 
 Fig. \ref{graph:third} by using the polynomial representations. 

We remark that at $\Delta=1$ our results reproduce the same
values as in \cite{Boos02,Sakai03}. It is immediate to see them from 
the following asymptotic expansions at $\eta=0$:
\begin{eqnarray}
 \zeta_\eta\left(1\right)&=&
-2\log 2 \eta^{-1}
+\frac{\pi^2}{6}\eta
-\frac{\pi^4}{180}\eta^3
+\frac{\pi^6}{945}\eta^5+\cdots
\\
 \zeta_\eta\left(3\right)&=&
\frac{3\zeta\left(3\right)}{2\pi^2}\eta^{-3}
-\frac{\pi^2}{30}\eta
+\frac{\pi^4}{189}\eta^3
-\frac{\pi^6}{450}\eta^5+\cdots
\\
 \zeta_\eta\left(5\right)&=&
-\frac{15\zeta\left(5\right)}{8\pi^4}\eta^{-5}
-\frac{\zeta\left(3\right)}{2\pi^2}\eta^{-3}
+\frac{31\pi^2}{1890}\eta
-\frac{41\pi^4}{11340}\eta^3
+\frac{31\pi^6}{14850}\eta^5+\cdots,\quad{\rm etc.}
\end{eqnarray}
\section{Conclusion}
We have analyzed multiple integral representations of correlation functions for
$XXZ$ model. In our results, we have given {\it polynomial representations } for all the 
 adjacent points correlation functions under the restriction that the number of the points is
 one to four. The {\it polynomial} means polynomial with respect to
 specific integrals (\ref{ea:notation_first_intro}) and (\ref{ea:notation_last_intro})
 where the coefficients are rational functions of $\sin \pi \eta$ and $\cos \pi \eta$. 
Here,  $\cos \pi \eta$ is equal to $\Delta$, and $\eta$ is either a real number ($-1 <\Delta \le 1$) 
or  a purely imaginary number (${1 < \Delta }$). It is intriguing that the correlation functions are given 
by the common expressions both in the massless regime and the massive regime.  Probably it will be possible 
to generalize our results to more general ${XYZ}$ model.

On the other hand, we can, in principle, study the correlation functions even further for ${n \ge 5}$ using 
our method. By surveying our results, we conjecture that there exists {\it polynomial representations} for 
all these correlation functions, where only the one-dimensional integrals  (\ref{ea:notation_first_intro}) and 
(\ref{ea:notation_last_intro}) with odd integer $j$ appear.

\section{Acknowledgements}
The authors are grateful to H. E. Boos, K. Hikami and M. Wadati for valuable comments and stimulating 
discussions. This work is  in part supported by Grant-in-Aid for the Scientific Research (B) No.~14340099 
and also by a 21st Century COE Program at Tokyo Tech``Nanometer-Scale Quantum Physics" from the Ministry 
of Education, Culture, Sports, Science and Technology. Further GK is supported by the JSPS research fellowships 
for young scientists and MS is supported by Grant-in Aid for Young Scientists (B) No. 14740228.

\appendix
\section{Canonical form for $ F\left[_{++++}^{++++}\right]$}
\label{sec:example}
In this section, we derive canonical form of $
F\left[_{++++}^{++++}\right]$
as an example. In the following, we assume $F^{(n)}_m$ as a polynomial with respect to
$y_1$, $y_2$, $y_3$ and $y_4$ including negative power terms.

The polynomial $T$  with respect to $
F\left[_{++++}^{++++}\right]$ are written as
\begin{eqnarray}
&&T\left(y_1,y_2,y_3,y_4\right)\nonumber\\&=& \frac{
\left(  y_1-1\right)^3
\left(  y_2-1\right)^2
\left(q y_2-1\right)
\left(  y_3-1\right)
\left(q y_3-1\right)^2
\left(q y_4-1\right)^3
}{
\left(y_1-q y_2\right)
\left(y_1-q y_3\right)
\left(y_1-q y_4\right)
\left(y_2-q y_3\right)
\left(y_2-q y_4\right)
\left(y_3-q y_4\right)
}
F_a^{(1)}
\nonumber\\
&=& 
\frac{
\left(  y_1-1\right)^3
\left(  y_2-1\right)^2
\left(q y_2-1\right)
\left(  y_3-1\right)
\left(q y_3-1\right)^2
\left(q y_4-1\right)^3
}{
\left(y_1-q y_2\right)
\left(y_1-q y_3\right)
\left(y_1-q y_4\right)
\left(y_2-q y_4\right)
\left(y_2-q y_3\right)
}
F_a^{(2)}
\nonumber\\&&{}
- \frac{
\left(  y_1-1\right)^3
\left(  y_2-1\right)^2
\left(q y_2-1\right)
\left(  y_3-1\right)
\left(q y_3-1\right)^2
\left(q y_4-1\right)^3
}{
\left(y_1-q y_2\right)
\left(y_1-q y_3\right)
\left(y_1-q y_4\right)
\left(y_3-q y_4\right)
\left(y_2-q y_3\right)
}
F_a^{(2)}
\nonumber\\&&{}
 \frac{
\left(  y_1-1\right)^3
\left(  y_2-1\right)^2
\left(q y_2-1\right)
\left(  y_3-1\right)
\left(q y_3-1\right)^2
\left(q y_4-1\right)^3
}{
\left(y_1-q y_2\right)
\left(y_1-q y_3\right)
\left(y_1-q y_4\right)
\left(y_3-q y_4\right)
\left(y_2-q y_4\right)
}
F_a^{(2)}.
\label{tmp:A_001}
\end{eqnarray}
Here the second equality is due to an elementary relation
\begin{eqnarray}
&&{} 
\frac1{
\left(y_k-y_l  q\right)
\left(y_j-y_l  q\right)
\left(y_j-y_k  q\right)}
\nonumber\\
&=&
\frac1{
\left(1- q\right)y_k}
\left[
\frac1{
\left(y_j-y_l  q\right)
\left(y_j-y_k  q\right)}
-
\frac1{
\left(y_k-y_l  q\right)
\left(y_j-y_k  q\right)}
+\frac1{
\left(y_k-y_l  q\right)
\left(y_j-y_l  q\right)
}
\right].
\nonumber\\
\label{eq:elementary_relation}
\end{eqnarray}
We named three terms in (\ref{tmp:A_001})
 as $I^{(2)}_1$, $I^{(2)}_2$ and $I^{(2)}_3$
order by order.
The two terms $I^{(2)}_1$ and $I^{(2)}_3$ are modified as
\begin{eqnarray}
 I_1^{(2)}&\sim&
 \frac{
\left(  y_1-1\right)^2
\left(  y_2-1\right)^2
\left(  y_3-1\right)
\left(q y_3-1\right)^2
\left(q y_4-1\right)^3
}{
\left(y_1-q y_2\right)
\left(y_1-q y_3\right)
\left(y_1-q y_4\right)
\left(y_2-q y_4\right)
}
F^{(3)}_1
\nonumber\\
&=&
\frac{
\left(  y_1-1\right)^2
\left(  y_2-1\right)^2
\left(  y_3-1\right)
\left(q y_3-1\right)^2
\left(q y_4-1\right)^3
}{
\left(y_1-q y_3\right)
\left(y_1-q y_2\right)
\left(y_1-q y_4\right)
}
F^{(4)}_1
\nonumber\\&&{}
-\frac{
\left(  y_1-1\right)^2
\left(  y_2-1\right)^2
\left(  y_3-1\right)
\left(q y_3-1\right)^2
\left(q y_4-1\right)^3
}{
\left(y_1-q y_3\right)
\left(y_1-q y_2\right)
\left(y_2-q y_4\right)
}
F^{(4)}_1
\nonumber\\&&{}
+\frac{
\left(  y_1-1\right)^2
\left(  y_2-1\right)^2
\left(  y_3-1\right)
\left(q y_3-1\right)^2
\left(q y_4-1\right)^3
}{
\left(y_1-q y_3\right)
\left(y_1-q y_4\right)
\left(y_2-q y_4\right)
}
F^{(4)}_1
\nonumber\\&=:&I^{(4)}_1+I^{(4)}_2+I^{(4)}_3
\\
I^{(2)}_3&\sim&
\frac{
\left(  y_1-1\right)^3
\left(  y_2-1\right)
\left(q y_2-1\right)
\left(  y_3-1\right)
\left(q y_3-1\right)
\left(q y_4-1\right)^3
}{
\left(y_1-q y_2\right)
\left(y_1-q y_3\right)
\left(y_1-q y_4\right)
\left(y_3-q y_4\right)
}
F^{(3)}_3
\nonumber\\
&=&
\frac{
\left(  y_1-1\right)^3
\left(  y_2-1\right)
\left(q y_2-1\right)
\left(  y_3-1\right)
\left(q y_3-1\right)
\left(q y_4-1\right)^3
}{
\left(y_1-q y_2\right)
\left(y_1-q y_4\right)
\left(y_3-q y_4\right)
}
F^{(4)}_3
\nonumber\\&&{}
-\frac{
\left(  y_1-1\right)^3
\left(  y_2-1\right)
\left(q y_2-1\right)
\left(  y_3-1\right)
\left(q y_3-1\right)
\left(q y_4-1\right)^3
}{
\left(y_1-q y_2\right)
\left(y_1-q y_3\right)
\left(y_3-q y_4\right)
}
F^{(4)}_3
\nonumber\\&&{}
+\frac{
\left(  y_1-1\right)^3
\left(  y_2-1\right)
\left(q y_2-1\right)
\left(  y_3-1\right)
\left(q y_3-1\right)
\left(q y_4-1\right)^3
}{
\left(y_1-q y_2\right)
\left(y_1-q y_3\right)
\left(y_1-q y_4\right)
}
F^{(4)}_3
\nonumber\\
&=:&I^{(4)}_{12}+I^{(4)}_{13}+I^{(4)}_{14}.
\end{eqnarray}
At the two {\it weak} equalities above, we used the relation
(\ref{eq:reduce_pole}).  We also used the elementary relation (\ref{eq:elementary_relation}) at the two 
equalities above. The term $I^{(2)}_2$ is modified as follows by use of elementary relations like (\ref{eq:elementary_relation}).
\begin{eqnarray}
I^{(2)}_2
&=&
\frac{
\left(  y_1-1\right)^3
\left(  y_2-1\right)^2
\left(q y_2-1\right)
\left(  y_3-1\right)
\left(q y_3-1\right)^2
\left(q y_4-1\right)^3
}{
\left(q-1\right)
\left(y_1-q y_2\right)
\left(y_1-q y_3\right)
\left(y_1-q y_4\right)
}
F^{(3)}_2
\nonumber\\&&{}
+\frac{
\left(  y_1-1\right)^3
\left(  y_2-1\right)^2
\left(q y_2-1\right)
\left(  y_3-1\right)
\left(q y_3-1\right)^2
\left(q y_4-1\right)^3
}{
\left(q-1\right)
\left(y_1-q y_3\right)
\left(y_1-q y_4\right)
\left(y_2-q y_3\right)
}
F^{(3)}_2
\nonumber\\&&{}
+\frac{
\left(  y_1-1\right)^3
\left(  y_2-1\right)^2
\left(q y_2-1\right)
\left(  y_3-1\right)
\left(q y_3-1\right)^2
\left(q y_4-1\right)^3
}{
\left(q^2-1\right)
\left(y_1-q y_2\right)
\left(y_1-q y_4\right)
\left(y_3-q y_4\right)
}
F^{(3)}_2
\nonumber\\&&{}
-\frac{
\left(  y_1-1\right)^3
\left(  y_2-1\right)^2
\left(q y_2-1\right)
\left(  y_3-1\right)
\left(q y_3-1\right)^2
\left(q y_4-1\right)^3
}{
\left(q^2-1\right)
\left(y_1-q y_2\right)
\left(y_1-q y_4\right)
\left(y_2-q y_3\right)
}
F^{(3)}_2
\nonumber\\&&{}
-\frac{
\left(  y_1-1\right)^3
\left(  y_2-1\right)^2
\left(q y_2-1\right)
\left(  y_3-1\right)
\left(q y_3-1\right)^2
\left(q y_4-1\right)^3
}{
\left(q-1\right)
\left(y_1-q y_2\right)
\left(y_1-q y_3\right)
\left(y_3-q y_4\right)
}
F^{(3)}_2
\nonumber\\&&{}
+\frac{
\left(  y_1-1\right)^3
\left(  y_2-1\right)^2
\left(q y_2-1\right)
\left(  y_3-1\right)
\left(q y_3-1\right)^2
\left(q y_4-1\right)^3
}{
\left(q^2-1\right)
\left(y_1-q y_2\right)
\left(y_2-q y_3\right)
\left(y_3-q y_4\right)
}
F^{(3)}_2
\nonumber\\&&{}
-\frac{
\left(  y_1-1\right)^3
\left(  y_2-1\right)^2
\left(q y_2-1\right)
\left(  y_3-1\right)
\left(q y_3-1\right)^2
\left(q y_4-1\right)^3
}{
\left(q-1\right)
\left(y_1-q y_3\right)
\left(y_2-q y_3\right)
\left(y_3-q y_4\right)
}
F^{(3)}_2
\nonumber\\&&{}
+\frac{
\left(  y_1-1\right)^3
\left(  y_2-1\right)^2
\left(q y_2-1\right)
\left(  y_3-1\right)
\left(q y_3-1\right)^2
\left(q y_4-1\right)^3
q}{
\left(q^2-1\right)
\left(y_1-q y_4\right)
\left(y_2-q y_3\right)
\left(y_3-q y_4\right)
}
F^{(3)}_2
\nonumber\\
&=:&
I^{(4)}_4+I^{(4)}_5+I^{(4)}_6+
I^{(4)}_7+I^{(4)}_8+I^{(4)}_9+
I^{(4)}_{10}+I^{(4)}_{11}.
\end{eqnarray}

It is rather troublesome to show the further evaluation directly. Then, we introduce graphs 
which indicate  certain sets of rational functions with respect to
$\{y_1,y_2,y_3,y_4\}$ as follows.

Each graph is made from three parts, i.e. circles, directed lines 
and non-directed lines. First, there are four circles in a graph, which are
placed at the four corners. 
The circle at upper left ( right ) represents the variable $y_1$ ( $y_2$ ), and 
the circle at lower left ( right ) represents the variable $y_3$ ( $y_4$ ). Next, each 
graph contains directed lines or non-directed lines which connects two circles and indicates 
a certain analytic property of rational functions. 
If a non-directed line connects $y_j$ and $y_k$,  it implies that the order of 
a pole at $y_j=y_k$ is $1$ at most. If $y_j$ and $y_k$ are connected by a 
directed line from $y_j$ to $y_k$, the order of a pole at $y_j=q y_k$ is $1$ at most, 
and at the same time the function has a factor $(y_j-1)^{m_j}(q y_k-1)^{m_k}$ with $m_j+m_k\geq 4$.
Finally, we assume the rational functions do not have any other pole except for the one 
 at $y_j=0$.  In this way we can define a set of rational functions corresponding to 
each graph.
For example,  $I^{(4)}_4$ is in the set
\begin{eqnarray}
\begin{psfrags}
  \psfig{file=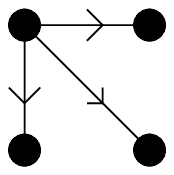 , scale = 0.9}
\end{psfrags}
.
\end{eqnarray} 

\begin{figure}
\begin{center}
\begin{psfrags}
  \psfig{file=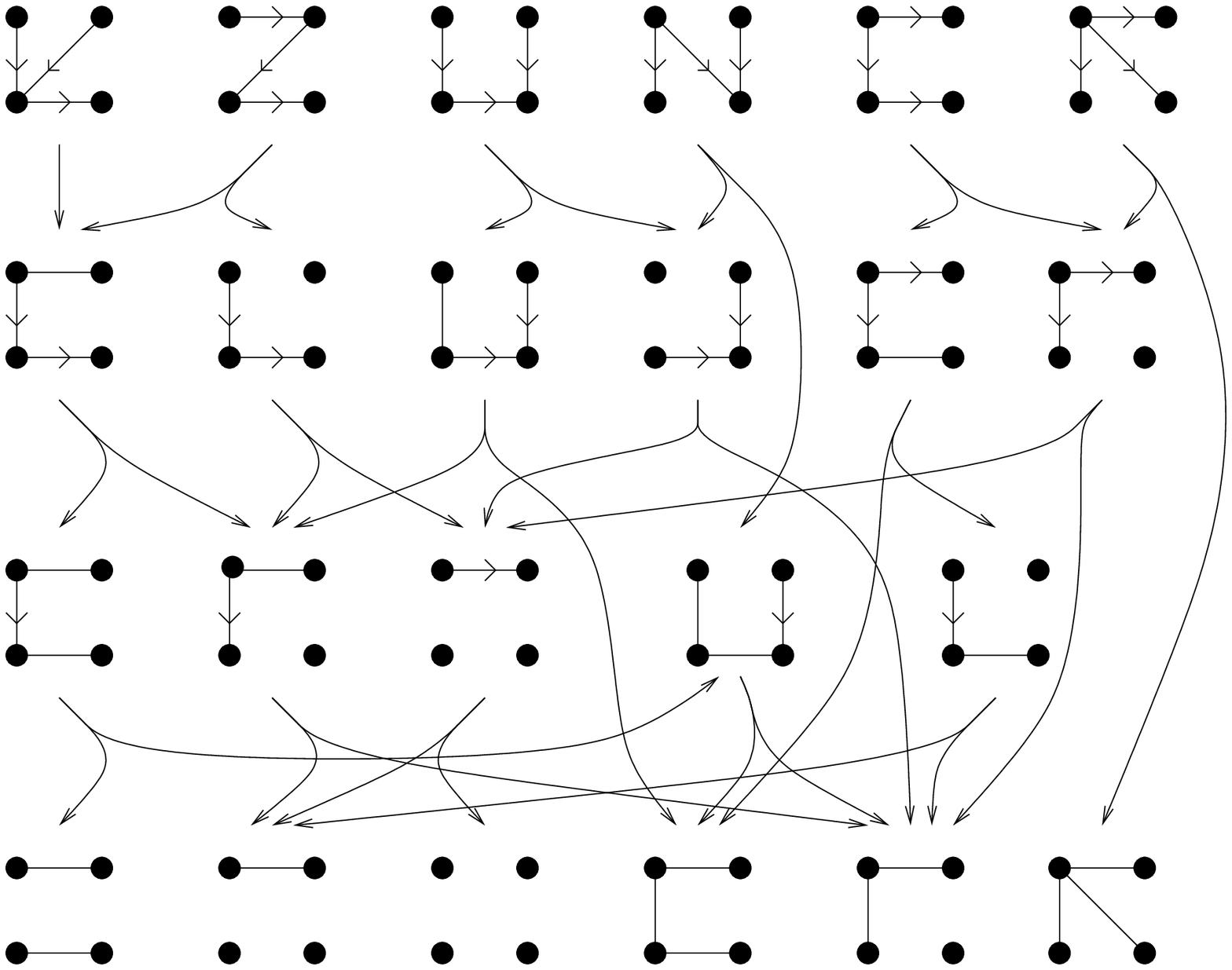 , scale = 0.9}
\end{psfrags}
\end{center}
\caption{The modification stream of rational functions 1}
\label{fig:modification_stream_1}
\end{figure}

\begin{figure}
\begin{center}
\begin{psfrags}
  \psfig{file=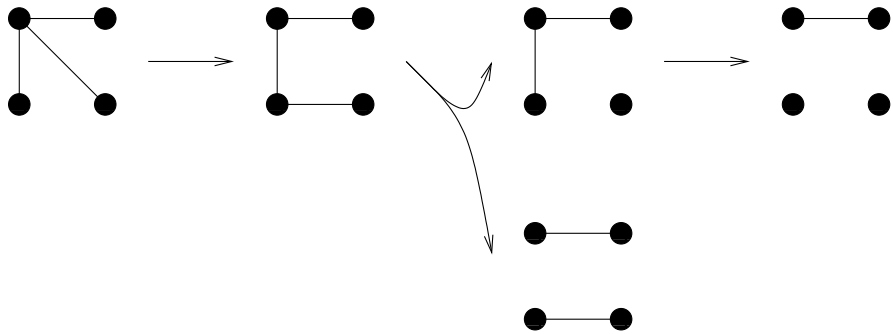 , scale = 0.9}
\end{psfrags}
\end{center}
\caption{The modification stream of rational functions 2}
\label{fig:modification_stream_2}
\end{figure}

We shall show the outline of further evaluation by means of these graphs.
In Fig. \ref{fig:modification_stream_1} and
\ref{fig:modification_stream_2}, some relations between the graphs are shown ;  
\begin{eqnarray}
\makebox{``graph1''}
&\longrightarrow&\makebox{ ``graph2''}
\end{eqnarray}
 indicates that any function in
``graph1'' can be modified into a sum of functions in ``graph2'' using the week equality.
Similarly, 
\begin{eqnarray}
\begin{array}{c}
\makebox{``graph1''}\\\\
\end{array}
&
\begin{psfrags}
  \psfig{file=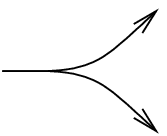 , scale = 0.6}
\end{psfrags}
&
\begin{array}{c}
\makebox{ ``graph2''} \\
\\
\makebox{ ``graph3''} \\\\
\end{array}
\end{eqnarray}
indicates that any function in ``graph1'' can be modified into a sum of
functions in ``graph2'' and ``graph3''. We can modify functions along the arrows between graphs 
in Fig. \ref{fig:modification_stream_1} and  \ref{fig:modification_stream_2}.
Below we prove it for some of the arrows. First, let us consider the upper left arrow in 
Fig. \ref{fig:modification_stream_1}
\begin{eqnarray}
\begin{psfrags}
  \psfig{file=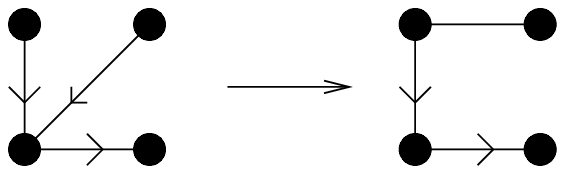 , scale = 0.9}.
\end{psfrags}
 \end{eqnarray}
This modification can be proved primitively as follows.
\begin{eqnarray}
&& \frac{
\left(  y_1-1\right)^{m_1}
\left(  y_2-1\right)^{m_2}
\left(  y_3-1\right)^{m_3}
\left(q y_3-1\right)^{m_4}
\left(q y_4-1\right)^{m_5}
}{
\left(y_1-q y_3\right)
\left(y_2-q y_3\right)
\left(y_3-q y_4\right)
}
F\left(y_1,y_2,y_3,y_4\right)
\nonumber\\&=& \frac{
\left(  y_1-1\right)^{m_1}
\left(  y_2-1\right)^{m_2}
\left(  y_3-1\right)^{m_3}
\left(q y_3-1\right)^{m_4}
\left(q y_4-1\right)^{m_5}
}{
\left(y_1- y_2\right)
\left(y_2-q y_3\right)
\left(y_3-q y_4\right)
}
F\left(y_1,y_2,y_3,y_4\right)
\nonumber\\&&{}+ \frac{
\left(  y_1-1\right)^{m_1}
\left(  y_2-1\right)^{m_2}
\left(  y_3-1\right)^{m_3}
\left(q y_3-1\right)^{m_4}
\left(q y_4-1\right)^{m_5}
}{
\left(y_1-q y_3\right)
\left(y_2- y_1\right)
\left(y_3-q y_4\right)
}
F\left(y_1,y_2,y_3,y_4\right)
\nonumber\\&\sim& \frac{
\left(  y_1-1\right)^{m_1}
\left(  y_2-1\right)^{m_2}
\left(  y_3-1\right)^{m_3}
\left(q y_3-1\right)^{m_4}
\left(q y_4-1\right)^{m_5}
}{
\left(y_1- y_2\right)
\left(y_2-q y_3\right)
\left(y_3-q y_4\right)
}
F\left(y_1,y_2,y_3,y_4\right)
\nonumber\\&&{}- \frac{
\left(  y_2-1\right)^{m_1}
\left(  y_1-1\right)^{m_2}
\left(  y_3-1\right)^{m_3}
\left(q y_3-1\right)^{m_4}
\left(q y_4-1\right)^{m_5}
}{
\left(y_2-q y_3\right)
\left(y_1- y_2\right)
\left(y_3-q y_4\right)
}
F\left(y_2,y_1,y_3,y_4\right), \nonumber \\ 
\end{eqnarray}
where $F$ is a polynomial including negative powers. 
At the weak equality, we used the relation (\ref{eq:transposition}).
Next, we prove the relation
\begin{eqnarray}
\begin{psfrags}
  \psfig{file=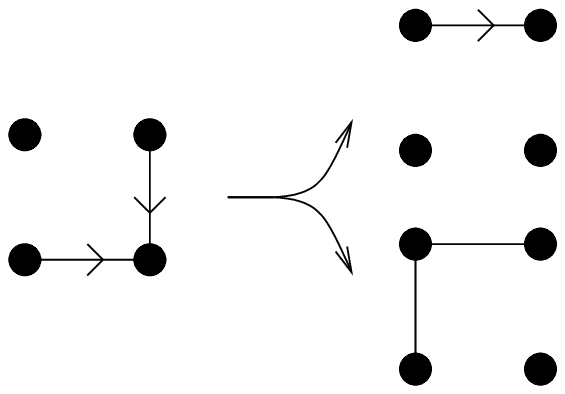 , scale = 0.9}
\end{psfrags} ,
\label{eq:expample_arrow_2}
\end{eqnarray}
which is one of arrows in Fig. \ref{fig:modification_stream_1}.
\begin{eqnarray}
&& \frac{
\left(  y_2-1\right)^{m_1}
\left(  y_3-1\right)^{m_2}
\left(q y_4-1\right)^{m_3}
}{
\left(y_2-q y_4\right)
\left(y_3-q y_4\right)
}
F\left(y_1,y_2,y_3,y_4\right)
\nonumber\\
&=&
 \frac{
\left(  y_2-1\right)^{m_1}
\left(q y_4-1\right)^{m_3}
}{
\left(y_2-q y_4\right)
}
\sum_{j=0}^{m_2-1} \left(  y_3-1\right)^{m_2-1-j}\left(q  y_4-1\right)^{j}
F\left(y_1,y_2,y_3,y_4\right)
\nonumber\\&&{}+
 \frac{
\left(  y_2-1\right)^{m_1}
\left(q y_4-1\right)^{m_2+m_3}
}{
\left(y_2-q y_4\right)
\left(y_3-q y_4\right)
}
F\left(y_1,y_2,y_3,y_4\right)
\nonumber\\ &\sim&
 \frac{
\left(  y_1-1\right)^{m_1}
\left(q y_2-1\right)^{m_3}
}{
\left(y_1-q y_2\right)
}
\sum_{j=0}^{m_2-1} \left(  y_3-1\right)^{m_2-1-j}\left(q  y_2-1\right)^{j}
F\left(y_4,y_1,y_3,y_2\right)
\nonumber\\&&{}+
 \frac{
\left(  y_2-1\right)^{m_1}
\left( y_1-1\right)^{m_2+m_3}
}{
\left(y_2- y_1\right)
\left(y_3- y_1\right)
}
F\left(y_4,y_2,y_3,q^{-1}y_1\right).
\end{eqnarray}
Here at the weak equality, we used the relation (\ref{eq:transposition}) and (\ref{eq:proposition1}).
The other arrows in  Fig. \ref{fig:modification_stream_1} are proved in  the same way as the case 
(\ref{eq:expample_arrow_2}). As for Fig. \ref{fig:modification_stream_2}, the left arrow is proved by 
the relation like (\ref{eq:elementary_relation}) and (\ref{eq:transposition}),
the middle arrow by (\ref{eq:reduce_pole_2}) and the right arrow  by (\ref{eq:reduce_pole}). Finally we note that 
any function in
\begin{eqnarray}
\begin{psfrags}
  \psfig{file=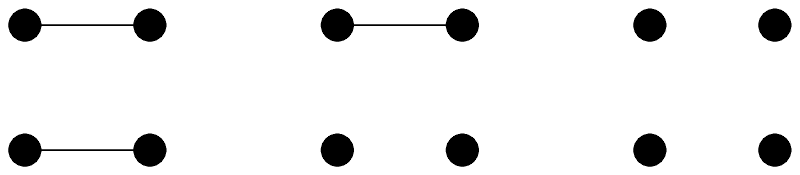 , scale = 0.9}
\end{psfrags} 
\end{eqnarray}
becomes canonical form using the relation (\ref{eq:corollary_1}) and
(\ref{eq:corollary_2}).

Now, let us  go back to modification of (\ref{tmp:A_001}).
We note that any
of functions  $I_1^{(4)}$,$I_2^{(4)}$,$\cdots$,$I_{14}^{(4)}$ is weekly equal to a function in
the sets in the Fig. \ref{fig:modification_stream_1}.
Then  the discussion above indicates that 
 $T(y_1,y_2,y_3,y_4)$ can be modified
into a canonical form according to the modification streams Fig.
\ref{fig:modification_stream_1} and \ref{fig:modification_stream_2} with the
relations (\ref{eq:corollary_1}) and
(\ref{eq:corollary_2}). 
Although we have actually come through the modifications, it is so tedious to show them here. 
Therefore, we show just the final result:
\begin{eqnarray}
&&
T\left(y_1,y_2,y_3,y_4\right)\nonumber\\&=& \frac{
\left(  y_1-1\right)^3
\left(  y_2-1\right)^2
\left(q y_2-1\right)
\left(  y_3-1\right)
\left(q y_3-1\right)^2
\left(q y_4-1\right)^3
}{
\left(y_1-q y_2\right)
\left(y_1-q y_3\right)
\left(y_1-q y_4\right)
\left(y_2-q y_3\right)
\left(y_2-q y_4\right)
\left(y_3-q y_4\right)
}
\nonumber\\&\sim&
\frac A{\left(y_2-y_1\right)\left(y_4-y_3\right)}
+\frac B{\left(y_2-y_1\right)}
+C,
\end{eqnarray}
where
\begin{eqnarray}
A&=&
 \frac{1-12 q+53 q^2-42 q^3+93 q^4+32 q^5+27 q^6+6 q^7+2 q^8}
      {128 q^4 \left( 1+q \right) \left(1+q+q^2 \right) }
\nonumber\\&&
-\frac{\left( 1+q \right)  \left( 1+3 q+q^2 \right)}{8 q^2}
        {y_3}
+\frac{\left( 4-9 q+45 q^2+16 q^3+36 q^4+21 q^5+7 q^6 \right)}
      {16 q^2 \left( 1+q \right)  \left( 1+q+q^2 \right) }
       {{y_3}}^2
\nonumber\\&&
-\frac{\left( 1-3 q+10 q^2+q^3+6 q^4+4 q^5+q^6 \right)}
      {16 q^2 \left(1+q+q^2 \right) }
       {{y_3}}^3
+\frac{\left( 3+22 q+70 q^2+22 q^3+3 q^4 \right)}{64 q^2}
        {y_1}{y_3}
\nonumber\\&&
-\frac{\left( 1+q \right)\left(1+2 q+24 q^2+2 q^3+q^4 \right)}
      {16 q^2}
        {y_1}{{y_3}}^2
+\frac{{\left( 1+q \right) }^2 \left( 1+18 q^2+q^4 \right)}
      {64 q^2}
        {y_1} {{y_3}}^3
\nonumber\\&&
+\frac{\left(2+134 q+91 q^2+512 q^3+413 q^4+214 q^5+53 q^6+20 q^7+q^8\right)}
      {128 q \left( 1+q \right)\left(1+q+q^2 \right) }
       {{y_1}}^2{{y_3}}^2
\nonumber\\&&
-\frac{\left( 7+24 q^2+19 q^3+6 q^4+3 q^5+q^6 \right)}
      {16 \left(1+q+q^2 \right)}
       {{y_1}}^2{{y_3}}^3
\nonumber\\&&
+\frac{{\left( 1+q \right) }^2 \left( 3-5 q+14 q^2-5 q^3+3 q^4 \right)}
      {64 \left(1+q+q^2 \right) }
       {{y_1}}^3 {{y_3}}^3,
\\
B&=&
 \frac{\left( 6-48 q+7 q^2+341 q^3-101 q^4-76 q^5-127 q^6+125 q^7+45 q^8 \right)}
      {64 q^3 \left( 1+q \right)  \left( 1+q+q^2 \right) }
        {y_4}
\nonumber\\&&
-\frac{\left( 2-25 q+58 q^2+78 q^3+33 q^4+77 q^5-104 q^6-14 q^7+81 q^8+18 q^9 \right)}
      {32 q^3 \left( 1+q \right)  \left( 1+q+q^2 \right) }
       {{y_4}}^2
\nonumber\\&&
+\frac{\left( 1-13 q+44 q^2+66 q^3-38 q^4+98 q^5+31 q^6-51 q^7+6 q^8+50 q^9+10 q^{10} \right)}
      {64 q^3 \left( 1+q \right)  \left( 1+q+q^2 \right) }
       {{y_4}}^3
\nonumber\\&&
+\frac{\left( -15+92 q-67 q^2+437 q^3-77 q^4+141 q^5-328 q^6+180 q^7+121 q^8 \right)}
      {64 q^2 \left( 1+q \right)  \left( 1+q+q^2 \right) }
        {y_3} {{y_4}}^2
\nonumber\\&&
-\frac{\left( -4+47 q-41 q^2+108 q^3-12 q^4+76 q^5-108 q^6+41 q^7+41 q^8 \right)}
      {64 q^2 \left( 1+q+q^2 \right) }
        {y_3} {{y_4}}^3
\nonumber\\&&
+\frac{\left( 7+q-4 q^2+44 q^3+5 q^4+13 q^5-24 q^6+14 q^7+11 q^8+3 q^9 \right)}
      {32 q \left( 1+q \right)  \left( 1+q+q^2 \right) }
       {{y_3}}^2 {{y_4}}^3
\nonumber\\&&
-\frac{\left( 1+q \right)}{4}
        {y_1} {y_4}
+\frac{\left( 2-q+2 q^2 \right)}{2}
        {y_1} {{y_4}}^2
-\frac{3 \left( 1+q \right)  \left( 1-q+q^2 \right)}{8}
        {y_1} {{y_4}}^3
\nonumber\\&&
-\frac{5 \left( 1+q \right)  \left( 1+q^2 \right)}{16}
        {y_1} {y_3} {{y_4}}^2
+\frac{5 {\left( 1+q \right) }^2 \left( 1-q+q^2 \right)}{32}
        {y_1} {y_3} {{y_4}}^3
\nonumber\\&&
-\frac{\left( 1+q \right)  \left( 1+q^2 \right)  \left( 1-q+q^2 \right)}{16}
        {y_1} {{y_3}}^2 {{y_4}}^3
\nonumber\\&&
-\frac{\left( -52+543 q^2-199 q^3-159 q^4-273 q^5+182 q^6+9 q^7 \right)}
      {64 q \left( 1+q \right)  \left(1+q+q^2 \right) }
       {{y_1}}^2 {y_4}
\nonumber\\&&
+\frac{\left( -53+181 q-15 q^2+96 q^3-q^4-429 q^5+197 q^6+60 q^7 \right)}
      {64 q \left( 1+q+q^2 \right) }
       {{y_1}}^2 {{y_4}}^2
\nonumber\\&&
-\frac{\left( -14+ 53 q+58 q^2-81 q^3+126 q^4-2 q^5-148 q^6-47 q^7+76 q^8+15 q^9 \right)}
      {64 q \left( 1+q \right)  \left( 1+q+q^2 \right) }
       {{y_1}}^2 {{y_4}}^3
\nonumber\\&&
-\frac{\left(-13+110 q-118 q^2+659 q^3-55 q^4+116 q^5-586 q^6+161 q^7+218 q^8 \right)}
      {64 q \left( 1+q \right)  \left( 1+q+q^2 \right) }
       {{y_1}}^2 {y_3} {{y_4}}^2
\nonumber\\&&
+\frac{\left( -3+55 q-62 q^2+168 q^3+28 q^4+100 q^5-177 q^6+25 q^7+70 q^8 \right)}
      {64 q \left( 1+q+q^2 \right) }
       {{y_1}}^2 {y_3} {{y_4}}^3
\nonumber\\&&
-\frac{\left( 16-q-14 q^2+149 q^3+49 q^4+50 q^5-79 q^6+18 q^7+30 q^8+10 q^9 \right)}
      {64 \left( 1+q \right)  \left( 1+q+q^2 \right) }
       {{y_1}}^2 {{y_3}}^2 {{y_4}}^3
\nonumber\\&&
+\frac{\left( -13-q+155 q^2-40 q^3-23 q^4-79 q^5+53 q^6+24 q^7 \right)}
      {64 q \left(1+q+q^2 \right) }
       {{y_1}}^3 {y_4}
\nonumber\\&&
-\frac{\left( 1+q \right)  \left( -14+51 q+9 q^2+37 q^3+18 q^4-117 q^5+61 q^6+15 q^7 \right)}
      {64 q \left( 1+q+q^2 \right) }
       {{y_1}}^3 {{y_4}}^2
\nonumber\\&&
+\frac{\left( -2+8 q+11 q^2-6 q^3+22 q^4+4 q^5-16 q^6-4 q^7+11 q^8+2 q^9 \right)}
      {32 q \left( 1+q+q^2 \right) }
       {{y_1}}^3 {{y_4}}^3
\nonumber\\&&
+\frac{\left( -4+31 q-23 q^2+199  q^3+18 q^4+59 q^5-143 q^6+47 q^7+60 q^8 \right)}
      {64 q \left( 1+q+q^2 \right) }
       {{y_1}}^3 {y_3} {{y_4}}^2
\nonumber\\&&
-\frac{\left( 1+q \right)  \left( -1+16 q-13 q^2+51 q^3+16 q^4+33 q^5-42 q^6+8 q^7+20 q^8 \right)}
      {64 q \left( 1+q+q^2 \right) }
       {{y_1}}^3 {y_3} {{y_4}}^3
\nonumber\\&&
+\frac{\left( 5+q-q^2+45 q^3+20 q^4+20 q^5-17 q^6+7 q^7+9 q^8+3 q^9 \right)}
      {64 \left( 1+q+q^2 \right) }
       {{y_1}}^3 {{y_3}}^2 {{y_4}}^3,
\\
C&=&
\frac{
 1+3 q-16 q^2-50 q^3+51 q^4-59 
    q^5+18 q^6-29 q^7+29 q^8-15 q^9-9 
    q^{10} }{64 
 q \left( 1+q \right) \left( 1+q+q^2 \right) } {y_2} {{y_3}}^2 {{y_4}}^3.
\nonumber\\
\end{eqnarray}
\section{Results}
\label{sec:results}
We use some notations defined in (\ref{ea:notation_first})$\sim$(\ref{ea:notation_last}).
\begin{eqnarray}
F\left[
\begin{array}{c}
+\\
+
\end{array}
\right]
&=&
\frac{1}{2}
\\
F\left[
\begin{array}{cc}
+&+\\
+&+
\end{array}
\right]
&=&\frac{1}{2}
-\frac{{c_1}  }{2  \pi   {s_1}}{\zeta _\eta}(1)
-\frac{1}{2  {\pi }^2}{\zeta'_\eta}(1)
\\
F\left[
\begin{array}{cc}
+&-\\
-&+
\end{array}
\right]
&=&
\frac{1}{2 \pi  {s_1}}{\zeta _\eta}(1)
+\frac{{c_1}}{2 {\pi }^2} {\zeta'_\eta}(1)
\\
F\left[
\begin{array}{ccc}
+&+&+\\
+&+&+
\end{array}
\right]
&=&\frac{1}{2}
-\frac{ 1 + 2 {c_2} }{2 \pi  {s_2}}{\zeta _\eta}(1)
-\frac{3 }{4 {\pi }^2} {\zeta'_\eta}(1)
+\frac{3 {s_1} }{8 \pi  {c_1}}{\zeta _\eta}(3)
+\frac{ 1 - {c_2} }{16 {\pi }^2}  {\zeta'_\eta}(3)
\\
 F\left[
\begin{array}{ccc}
+&+&-\\
-&+&+
\end{array}
\right]
&=&
\frac{1}{2 \pi  {s_2}}{\zeta _\eta}(1)
+\frac{{c_2} }{4 {\pi }^2} {\zeta'_\eta}(1)
-\frac{3 {c_2} \left( 1 - {c_2} \right)  }{8 \pi  {s_2}} {\zeta _\eta}(3)
-\frac{1-{c_2} }{16 {\pi }^2}{\zeta'_\eta}(3)
\end{eqnarray}

\begin{eqnarray}
&& F\left[
\begin{array}{cccc}
+&+&+&+\\
+&+&+&+
\end{array}
\right]
\nonumber\\
&=&
\frac{1}{2}
-\frac{{c_2} \left( 5+6 {c_2} \right) }{4 \pi {c_1} {s_3}}{\zeta _\eta}(1)
-\frac{11 }{12 {\pi }^2}{\zeta'_\eta}(1)
-\frac{21+61{c_2}+ 38 {c_4}+3 {c_6}+{c_8}}{32 \pi {c_1} {s_3}}{\zeta _\eta}(3)
\nonumber\\&&
-\frac{5+6 {c_2}+3 {c_4}}{48 {\pi }^2} {\zeta'_\eta}(3)
-\frac{5 \left( 33+61 {c_2}+22 {c_4}+3 {c_6}+{c_8} \right) }{64 \pi {c_1} {s_3}}{\zeta _\eta}(5)
-\frac{7+2 {c_2}+{c_4}}{16 {\pi }^2}{\zeta'_\eta}(5)
\nonumber\\&&
+\frac{109+127 {c_2}+58 {c_4}+5 {c_6}+{c_8}}{32 {\pi }^2 {s_1} {s_3}}{\zeta _\eta}(1) {\zeta _\eta}(3)
+\frac{{c_1} \left( 33+16 {c_2}+{c_4} \right) }{16 {\pi }^3 {s_1}} {\zeta'_\eta}(1){\zeta _\eta}(3) 
\nonumber\\&&
+\frac{{c_1} \left( 9+{c_4} \right) }{16 {\pi }^3 {s_1}} {\zeta _\eta}(1) {\zeta'_\eta}(3)
+\frac{35+22 {c_2}+3 {c_4}}{96 {\pi }^4}{\zeta'_\eta}(1) {\zeta'_\eta}(3)
\nonumber\\&&
+\frac{5 \left( 121+163 {c_2}+70 {c_4}+5 {c_6}+{c_8} \right) }{64 {\pi }^2 {s_1} {s_3}}{\zeta _\eta}(1) {\zeta _\eta}(5)
+\frac{5 {c_1} \left( 21+8 {c_2}+{c_4} \right) }{16 {\pi }^3 {s_1}}  {\zeta'_\eta}(1){\zeta _\eta}(5)
\nonumber\\&&
+\frac{{c_1} \left( 21+8 {c_2}+{c_4} \right) }{16 {\pi }^3 {s_1}}{\zeta _\eta}(1) {\zeta'_\eta}(5)
+\frac{ 35+22 {c_2}+3 {c_4}  }{32 {\pi }^4}{\zeta'_\eta}(1) {\zeta'_\eta}(5)
\nonumber\\&&
-\frac{3 \left( 121+163 {c_2}+70 {c_4}+5 {c_6}+{c_8} \right) }{128 {\pi }^2 {s_1} {s_3}}{{\zeta _\eta}(3)}^2
-\frac{{c_1} \left( 21+8 {c_2}+{c_4} \right) }{16 {\pi }^3 {s_1}}{\zeta _\eta}(3) {\zeta'_\eta}(3)
\nonumber\\&&
-\frac{35+22 {c_2}+3 {c_4}}{192 {\pi }^4}{{\zeta'_\eta}(3)}^2
\\
&& F\left[
\begin{array}{cccc}
+&-&+&-\\
+&-&+&-
\end{array}
\right]
\nonumber\\
&=&
 \frac{{c_2} }{4 \pi {c_1} {s_3}}
  {\zeta _\eta}(1)
+\frac{1}{12 {\pi }^2}
  {\zeta'_\eta}(1)
-\frac{ 24+23 {c_2}+14 {c_4}+{c_6} }{16 \pi{c_1} {s_3}}
  {\zeta _\eta}(3)
\nonumber\\&&
-\frac{ 4+3 {c_2} }{24 {\pi }^2}
  {\zeta'_\eta}(3)
-\frac{5 \left( 20+27 {c_2}+12 {c_4}+{c_6} \right) }{32\pi {c_1} {s_3}}
  {\zeta _\eta}(5)
-\frac{ 3+2 {c_2} }{8 {\pi }^2}
  {\zeta'_\eta}(5)
\nonumber\\&&
+\frac{3 \left( 4+{c_2} \right) \left( 3+6 {c_2}+{c_4} \right) }{16 {\pi }^2 {s_1} {s_3}}
  {\zeta _\eta}(1){\zeta _\eta}(3)
+\frac{5 {c_1} \left( 4+{c_2} \right)  }{8{\pi }^3 {s_1}}
  {\zeta'_\eta}(1){\zeta _\eta}(3)
\nonumber\\&&
+\frac{{c_1} \left( 4+{c_2} \right) }{8{\pi }^3 {s_1}}
  {\zeta _\eta}(1){\zeta'_\eta}(3)
+\frac{ 19+11 {c_2} }{48 {\pi }^4}
  {\zeta'_\eta}(1){\zeta'_\eta}(3)
\nonumber\\&&
+\frac{15 \left( 38+63 {c_2}+18 {c_4}+{c_6} \right) }{64 {\pi }^2 {s_1} {s_3}}
  {\zeta _\eta}(1){\zeta _\eta}(5)
+\frac{15 {c_1} \left( 4+{c_2} \right)}{8{\pi }^3 {s_1}}
  {\zeta'_\eta}(1){\zeta _\eta}(5)
\nonumber\\&&
+\frac{3 {c_1} \left( 4+{c_2} \right) }{8 {\pi }^3 {s_1}}
  {\zeta _\eta}(1){\zeta'_\eta}(5)
+\frac{ 19+11 {c_2} }{16 {\pi }^4}
  {\zeta'_\eta}(1){\zeta'_\eta}(5)
\nonumber\\&&
-\frac{9 \left( 38+63 {c_2}+18 {c_4}+{c_6} \right) }{128 {\pi }^2 {s_1} {s_3}}
  {{\zeta _\eta}(3)}^2
-\frac{3 {c_1} \left( 4+{c_2} \right)}{8{\pi }^3 {s_1}}
  {\zeta _\eta}(3){\zeta'_\eta}(3)
\nonumber\\&&
-\frac{ 19+11 {c_2} }{96 {\pi }^4}
  {{\zeta'_\eta}(3)}^2
\\
&& F\left[
\begin{array}{cccc}
+&+&+&-\\
+&+&-&+
\end{array}
\right]
\nonumber\\
&=&
 \frac{{c_2} }{2 \pi {s_3}}
  {\zeta _\eta}(1)
+\frac{{c_1} }{6 {\pi }^2}
  {\zeta'_\eta}(1)
+\frac{ 11+41 {c_2}+9 {c_4}+{c_6} }{16 \pi {s_3}}
  {\zeta _\eta}(3)
\nonumber\\&&
+\frac{{c_1} \left( 4+3 {c_2} \right) }{24 {\pi }^2}
  {\zeta'_\eta}(3)
+\frac{5 \left( 17+33 {c_2}+9 {c_4}+{c_6} \right)}{32 \pi {s_3}}
  {\zeta _\eta}(5)
+\frac{{c_1} \left( 7+3 {c_2} \right) }{16 {\pi }^2}
  {\zeta'_\eta}(5)
\nonumber\\&&
-\frac{\left( 4+{c_2} \right) \left( 20+27 {c_2}+12 {c_4}+{c_6} \right) }{16 {\pi }^2 {s_2} {s_3}}
  {\zeta _\eta}(1) {\zeta _\eta}(3)
-\frac{ 49+48 {c_2}+3 {c_4} }{32 {\pi }^3 {s_1}}
  {\zeta'_\eta}(1){\zeta _\eta}(3)
\nonumber\\&&
-\frac{ 11+8 {c_2}+{c_4} }{32 {\pi }^3 {s_1}}
  {\zeta _\eta}(1) {\zeta'_\eta}(3)
-\frac{{c_1} \left( 11+4 {c_2} \right) }{24 {\pi }^4}
 {\zeta'_\eta}(1) {\zeta'_\eta}(3)
\nonumber\\&&
-\frac{5 \left( 219+328 {c_2}+148 {c_4}+24 {c_6}+{c_8} \right) }{64 {\pi }^2 {s_2} {s_3}}
  {\zeta _\eta}(1) {\zeta _\eta}(5)
-\frac{5 \left( 15+14 {c_2}+{c_4} \right)}{16 {\pi }^3 {s_1}}
  {\zeta'_\eta}(1){\zeta _\eta}(5) 
\nonumber\\&&
-\frac{ 15+14 {c_2}+{c_4} }{16 {\pi }^3 {s_1}}
  {\zeta _\eta}(1) {\zeta'_\eta}(5)
-\frac{{c_1} \left( 11+4 {c_2} \right) }{8 {\pi }^4}
  {\zeta'_\eta}(1) {\zeta'_\eta}(5)
\nonumber\\&&
+\frac{3 \left( 219+328 {c_2}+148 {c_4}+24 {c_6}+{c_8} \right) }{128 {\pi }^2 {s_2} {s_3}}
 {{\zeta _\eta}(3)}^2
+\frac{ 15+14 {c_2}+{c_4} }{16 {\pi }^3 {s_1}}
  {\zeta _\eta}(3) {\zeta'_\eta}(3)
\nonumber\\&&
+\frac{{c_1} \left( 11+4 {c_2} \right) }{48 {\pi }^4}
 {{\zeta'_\eta}(3)}^2
\\
&& F\left[
\begin{array}{cccc}
+&+&+&-\\
+&-&+&+
\end{array}
\right]
\nonumber\\
&=&
 \frac{{c_1} }{2 \pi {s_3}}
  {\zeta _\eta}(1)
+\frac{{c_2} }{6 {\pi }^2}
  {\zeta'_\eta}(1)
+\frac{{c_1} \left( 19+3 {c_2}+9 {c_4} \right) }{8 \pi {s_3}}
  {\zeta _\eta}(3)
\nonumber\\&&
+\frac{ 3+11 {c_2} }{48 {\pi }^2}
  {\zeta'_\eta}(3)
+\frac{15 {c_1} \left( 2+2 {c_2}+{c_4} \right) }{8 \pi {s_3}}
  {\zeta _\eta}(5)
+\frac{{{c_1}}^2 \left( 4+{c_2} \right)}{8 {\pi }^2}
  {\zeta'_\eta}(5)
\nonumber\\&&
-\frac{3 \left( 28+51 {c_2}+18 {c_4}+3 {c_6} \right) }{32 {\pi }^2 {s_1} {s_3}}
  {\zeta _\eta}(1) {\zeta _\eta}(3)
-\frac{{c_1} \left( 9+{c_2} \right) \left( 3+2 {c_2} \right)}{16 {\pi }^3 {s_1}}
  {\zeta'_\eta}(1) {\zeta _\eta}(3)
\nonumber\\&&
-\frac{5 {{c_1}}^3 }{8 {\pi }^3 {s_1}}
  {\zeta _\eta}(1) {\zeta'_\eta}(3)
-\frac{5 {{c_1}}^2 \left( 5+{c_2} \right) }{48 {\pi }^4}
  {\zeta'_\eta}(1) {\zeta'_\eta}(3)
\nonumber\\&&
-\frac{5 \left( 106+177 {c_2}+66 {c_4}+11 {c_6} \right)  }{64 {\pi }^2 {s_1} {s_3}}
  {\zeta _\eta}(1) {\zeta _\eta}(5)
-\frac{5 {c_1} \left( 33+26 {c_2}+{c_4} \right)}{32 {\pi }^3 {s_1}}
  {\zeta'_\eta}(1) {\zeta _\eta}(5) 
\nonumber\\&&
-\frac{{c_1} \left( 33+26 {c_2}+{c_4} \right)}{32 {\pi }^3 {s_1}}
  {\zeta _\eta}(1) {\zeta'_\eta}(5)
-\frac{5 {{c_1}}^2 \left( 5+{c_2} \right)}{16 {\pi }^4}
  {\zeta'_\eta}(1) {\zeta'_\eta}(5)
\nonumber\\&&
+\frac{3 \left( 106+177 {c_2}+66 {c_4}+11 {c_6} \right) }{128 {\pi }^2 {s_1} {s_3}}
 {{\zeta _\eta}(3)}^2
+\frac{{c_1} \left( 33+26 {c_2}+{c_4} \right)}{32 {\pi }^3 {s_1}}
  {\zeta _\eta}(3) {\zeta'_\eta}(3)
\nonumber\\&&
+\frac{5 {{c_1}}^2 \left( 5+{c_2} \right) }{96 {\pi }^4}
 {{\zeta'_\eta}(3)}^2
\\
&& F\left[
\begin{array}{cccc}
+&+&-&+\\
+&-&+&+
\end{array}
\right]
\nonumber\\
&=&
 \frac{1}{2 \pi {s_3}}
  {\zeta _\eta}(1)
+\frac{{c_1} }{6 {\pi }^2}
  {\zeta'_\eta}(1)
+\frac{ 19+5 {c_2}+7 {c_4} }{8 \pi {s_3}}
  {\zeta _\eta}(3)
\nonumber\\&&
+\frac{7 {c_1} }{24 {\pi }^2}
  {\zeta'_\eta}(3)
+\frac{15 \left( 2+2 {c_2}+{c_4} \right) }{8 \pi {s_3}}
  {\zeta _\eta}(5)
+\frac{{c_1} \left( 4+{c_2} \right) }{8 {\pi }^2}
  {\zeta'_\eta}(5)
\nonumber\\&&
-\frac{3 \left( 28+51 {c_2}+18 {c_4}+3 {c_6} \right)}{16 {\pi }^2 {s_2} {s_3}}
  {\zeta _\eta}(1) {\zeta _\eta}(3)
-\frac{\left( 9+{c_2} \right) \left( 3+2 {c_2} \right)  } {16 {\pi }^3 {s_1}}
  {\zeta'_\eta}(1) {\zeta _\eta}(3)
\nonumber\\&&
-\frac{5 {{c_1}}^2 }{8 {\pi }^3 {s_1}}
  {\zeta _\eta}(1) {\zeta'_\eta}(3)
-\frac{5 {c_1} \left( 5+{c_2} \right) }{48 {\pi }^4}
  {\zeta'_\eta}(1) {\zeta'_\eta}(3)
\nonumber\\&&
-\frac{5 \left( 106+177 {c_2}+66 {c_4}+11 {c_6} \right)}{32 {\pi }^2 {s_2} {s_3}}
  {\zeta _\eta}(1) {\zeta _\eta}(5)
-\frac{5 \left( 33+26 {c_2}+{c_4} \right)  }{32 {\pi }^3 {s_1}}
  {\zeta'_\eta}(1) {\zeta _\eta}(5)
\nonumber\\&&
-\frac{ 33+26 {c_2}+{c_4} }{32 {\pi }^3 {s_1}}
  {\zeta _\eta}(1) {\zeta'_\eta}(5)
-\frac{5 {c_1} \left( 5+{c_2} \right) }{16 {\pi }^4}
  {\zeta'_\eta}(1) {\zeta'_\eta}(5)
\nonumber\\&&
+\frac{3 \left( 106+177 {c_2}+66 {c_4}+11 {c_6} \right) }{64 {\pi }^2 {s_2} {s_3}}
 {{\zeta _\eta}(3)}^2
+\frac{ 33+26 {c_2}+{c_4} }{32 {\pi }^3 {s_1}}
  {\zeta _\eta}(3) {\zeta'_\eta}(3)
\nonumber\\&&
+\frac{5 {c_1} \left( 5+{c_2} \right)}{96 {\pi }^4}
 {{\zeta'_\eta}(3)}^2
\\
&& F\left[
\begin{array}{cccc}
+&+&+&-\\
-&+&+&+
\end{array}
\right]
\nonumber\\
&=&
 \frac{1}{2 \pi {s_3}}
  {\zeta _\eta}(1)
-\frac{{c_1} \left( 1-2 {c_2} \right) }{6 {\pi }^2}
  {\zeta'_\eta}(1)
+\frac{ 15+7 {c_2}+5 {c_4}+4 {c_6} }{8 \pi {s_3}}
  {\zeta _\eta}(3)
\nonumber\\&&
-\frac{{c_1} \left( 1-8 {c_2} \right) }{24 {\pi }^2}
  {\zeta'_\eta}(3)
+\frac{5 \left( 5+6 {c_2}+3 {c_4}+{c_6} \right) }{8 \pi {s_3}}
  {\zeta _\eta}(5)
+\frac{{c_1} \left( 5+4 {c_2}+{c_4} \right) }{16 {\pi }^2}
  {\zeta'_\eta}(5)
\nonumber\\&&
-\frac{{c_1} \left( 14+48 {c_2}+9 {c_4}+4 {c_6} \right) }{8 {\pi }^2 {s_1} {s_3}}
  {\zeta _\eta}(1) {\zeta _\eta}(3)
-\frac{ 34+53 {c_2}+12 {c_4}+{c_6} }{32 {\pi }^3 {s_1}}
  {\zeta'_\eta}(1) {\zeta _\eta}(3)
\nonumber\\&&
-\frac{ 5+3 {c_2}+2 {c_4} }{16 {\pi }^3 {s_1}}
  {\zeta _\eta}(1) {\zeta'_\eta}(3)
-\frac{{{c_1}}^3 \left( 11+4 {c_2} \right)}{24 {\pi }^4}
  {\zeta'_\eta}(1) {\zeta'_\eta}(3)
\nonumber\\&&
-\frac{5 {c_1} \left( 23+48 {c_2}+15 {c_4}+4 {c_6} \right) }{16 {\pi }^2 {s_1} {s_3}}
  {\zeta _\eta}(1) {\zeta _\eta}(5)
-\frac{5 {{c_1}}^2 \left( 15+14 {c_2}+{c_4} \right)}{16 {\pi }^3 {s_1}}
  {\zeta'_\eta}(1) {\zeta _\eta}(5)
\nonumber\\&&
-\frac{{{c_1}}^2 \left( 15+14 {c_2}+{c_4} \right) }{16 {\pi }^3 {s_1}}
  {\zeta _\eta}(1) {\zeta'_\eta}(5)
-\frac{{{c_1}}^3 \left( 11+4 {c_2} \right) }{8 {\pi }^4}
  {\zeta'_\eta}(1) {\zeta'_\eta}(5)
\nonumber\\&&
+\frac{3 {c_1} \left( 23+48 {c_2}+15 {c_4}+4 {c_6} \right) }{32 {\pi }^2 {s_1} {s_3}}
 {{\zeta _\eta}(3)}^2
+\frac{{{c_1}}^2 \left( 15+14 {c_2}+{c_4} \right) }{16 {\pi }^3 {s_1}}
  {\zeta _\eta}(3) {\zeta'_\eta}(3)
\nonumber\\&&
+\frac{{{c_1}}^3 \left( 11+4 {c_2} \right) }{48 {\pi }^4}
 {{\zeta'_\eta}(3)}^2
\\
&&F\left[
\begin{array}{cccc}
+&+&-&-\\
-&+&-&+
\end{array}
\right]
\nonumber\\
&=&
 \frac{1}{4 \pi {s_3}}
  {\zeta _\eta}(1)
-\frac{{c_1} \left(  1-2 {c_2} \right) }{12 {\pi }^2}
  {\zeta'_\eta}(1)
-\frac{{c_2} \left( 19+12 {c_2} \right) }{8 \pi {s_3}}
  {\zeta _\eta}(3)
\nonumber\\&&
-\frac{{c_1} \left( 5+2 {c_2} \right)}{24 {\pi }^2}
  {\zeta'_\eta}(3)
-\frac{5 \left( 8+15 {c_2}+6 {c_4}+{c_6} \right) }{16 \pi {s_3}}
  {\zeta _\eta}(5)
-\frac{{c_1} \left( 2+3 {c_2} \right) }{8 {\pi }^2}
  {\zeta'_\eta}(5)
\nonumber\\&&
+\frac{{c_1} \left( 56+57 {c_2}+36 {c_4}+{c_6} \right) }{16 {\pi }^2 {s_1} {s_3}}
  {\zeta _\eta}(1) {\zeta _\eta}(3)
+\frac{ 43+46 {c_2}+11 {c_4} }{32 {\pi }^3 {s_1}}
  {\zeta'_\eta}(1) {\zeta _\eta}(3)
\nonumber\\&&
+\frac{ 5+14 {c_2}+{c_4} }{32 {\pi }^3 {s_1}}
  {\zeta _\eta}(1) {\zeta'_\eta}(3)
+\frac{{{c_1}}^3 \left( 14+{c_2} \right)}{24 {\pi }^4}
  {\zeta'_\eta}(1) {\zeta'_\eta}(3)
\nonumber\\&&
+\frac{15 {c_1} \left( 20+27 {c_2}+12 {c_4}+{c_6} \right) }{32 {\pi }^2 {s_1} {s_3}}
  {\zeta _\eta}(1) {\zeta _\eta}(5)
+\frac{15 {{c_1}}^2 \left( 3+2 {c_2} \right)  }{8 {\pi }^3 {s_1}}
  {\zeta'_\eta}(1) {\zeta _\eta}(5)
\nonumber\\&&
+\frac{3 {{c_1}}^2 \left( 3+2 {c_2} \right)}{8 {\pi }^3 {s_1}}
  {\zeta _\eta}(1) {\zeta'_\eta}(5)
+\frac{{{c_1}}^3 \left( 14+{c_2} \right)}{8 {\pi }^4}
  {\zeta'_\eta}(1) {\zeta'_\eta}(5)
\nonumber\\&&
-\frac{9 {c_1} \left( 20+27 {c_2}+12 {c_4}+{c_6} \right) }{64 {\pi }^2 {s_1} {s_3}}
 {{\zeta _\eta}(3)}^2
-\frac{3 {{c_1}}^2 \left( 3+2 {c_2} \right) }{8 {\pi }^3 {s_1}}
  {\zeta _\eta}(3) {\zeta'_\eta}(3)
\nonumber\\&&
-\frac{{{c_1}}^3 \left( 14+{c_2} \right)}{48 {\pi }^4}
 {{\zeta'_\eta}(3)}^2
\\
&& F\left[
\begin{array}{cccc}
+&+&-&-\\
-&-&+&+
\end{array}
\right]
\nonumber\\
&=&
 \frac{1}{4 \pi {c_1} {s_3}}
  {\zeta _\eta}(1)
+\frac{{c_4}}{12 {\pi }^2}
  {\zeta'_\eta}(1)
-\frac{ 6+15 {c_2}+10 {c_4} }{8 \pi {c_1} {s_3}}
  {\zeta _\eta}(3)
\nonumber\\&&
-\frac{ 6+{c_4} }{24 {\pi }^2}
  {\zeta'_\eta}(3)
-\frac{5 {c_2} \left( 11+15 {c_2}+3 {c_4}+{c_6} \right) }{16 \pi {c_1} {s_3}}
  {\zeta _\eta}(5)
-\frac{ 2+2 {c_2}+{c_4} }{8 {\pi }^2}
  {\zeta'_\eta}(5)
\nonumber\\&&
+\frac{ 46+57 {c_2}+36 {c_4}+11 {c_6} }{16 {\pi }^2 {s_1} {s_3}}
  {\zeta _\eta}(1) {\zeta _\eta}(3)
+\frac{{c_1} \left( 12+9 {c_2}+4 {c_4} \right) }{8 {\pi }^3 {s_1}}
  {\zeta'_\eta}(1) {\zeta _\eta}(3)
\nonumber\\&&
+\frac{5 {c_1} {c_2} }{8 {\pi }^3 {s_1}}
  {\zeta _\eta}(1) {\zeta'_\eta}(3)
+\frac{{{c_1}}^2 \left( 13+16 {c_2}+{c_4} \right)}{48 {\pi }^4}
  {\zeta'_\eta}(1) {\zeta'_\eta}(3)
\nonumber\\&&
+\frac{5 \left( 51+76 {c_2}+40 {c_4}+12 {c_6}+{c_8} \right) }{32 {\pi }^2 {s_1} {s_3}}
  {\zeta _\eta}(1) {\zeta _\eta}(5)
+\frac{5 {c_1} \left( 6+7 {c_2}+2 {c_4} \right) }{8 {\pi }^3 {s_1}}
  {\zeta'_\eta}(1) {\zeta _\eta}(5) 
\nonumber\\&&
+\frac{{c_1} \left( 6+7 {c_2}+2 {c_4} \right)}{8 {\pi }^3 {s_1}}
  {\zeta _\eta}(1) {\zeta'_\eta}(5)
+\frac{{{c_1}}^2 \left( 13+16 {c_2}+{c_4} \right)}{16 {\pi }^4}
  {\zeta'_\eta}(1) {\zeta'_\eta}(5)
\nonumber\\&&
-\frac{3 \left( 51+76 {c_2}+40 {c_4}+12 {c_6}+{c_8} \right) }{64 {\pi }^2 {s_1} {s_3}}
 {{\zeta _\eta}(3)}^2
-\frac{{c_1} \left( 6+7 {c_2}+2 {c_4} \right) }{8 {\pi }^3 {s_1}}
  {\zeta _\eta}(3) {\zeta'_\eta}(3)
\nonumber\\&&
-\frac{{{c_1}}^2 \left( 13+16 {c_2}+{c_4} \right)}{96 {\pi }^4}
 {{\zeta'_\eta}(3)}^2
\\
&& F\left[
\begin{array}{cccc}
+&-&+&-\\
-&+&-&+
\end{array}
\right]
\nonumber\\
&=&
 \frac{1}{4 \pi {c_1} {s_3}}
  {\zeta _\eta}(1)
+\frac{{c_2}}{12 {\pi }^2}
  {\zeta'_\eta}(1)
-\frac{ 11+37 {c_2}+13 {c_4}+{c_6} }{16 \pi {c_1} {s_3}}
  {\zeta _\eta}(3)
\nonumber\\&&
-\frac{ 3+4 {c_2} }{24 {\pi }^2}
  {\zeta'_\eta}(3)
-\frac{5 \left( 8+15 {c_2}+6 {c_4}+{c_6} \right)}{16 \pi {c_1} {s_3}}
  {\zeta _\eta}(5)
-\frac{ 2+3 {c_2} }{8 {\pi }^2}
  {\zeta'_\eta}(5)
\nonumber\\&&
+\frac{ 55+60 {c_2}+33 {c_4}+2 {c_6} }{16 {\pi }^2 {s_1} {s_3}}
  {\zeta _\eta}(1) {\zeta _\eta}(3)
+\frac{5 {c_1} \left( 3+2 {c_2} \right) }{8 {\pi }^3 {s_1}}
  {\zeta'_\eta}(1) {\zeta _\eta}(3)
\nonumber\\&&
+\frac{{c_1} \left( 3+2 {c_2} \right) }{8 {\pi }^3 {s_1}}
  {\zeta _\eta}(1) {\zeta'_\eta}(3)
+\frac{{{c_1}}^2 \left( 14+{c_2} \right) }{24 {\pi }^4}
  {\zeta'_\eta}(1) {\zeta'_\eta}(3)
\nonumber\\&&
+\frac{15 \left( 20+27 {c_2}+12 {c_4}+{c_6} \right) }{32 {\pi }^2 {s_1} {s_3}}
  {\zeta _\eta}(1) {\zeta _\eta}(5)
+\frac{15 {c_1} \left( 3+2 {c_2} \right) }{8 {\pi }^3 {s_1}}
  {\zeta'_\eta}(1) {\zeta _\eta}(5)
\nonumber\\&&
+\frac{3 {c_1} \left( 3+2 {c_2} \right) }{8 {\pi }^3 {s_1}}
  {\zeta _\eta}(1) {\zeta'_\eta}(5)
+\frac{{{c_1}}^2 \left( 14+{c_2} \right)}{8 {\pi }^4}
  {\zeta'_\eta}(1) {\zeta'_\eta}(5)
\nonumber\\&&
-\frac{9 \left( 20+27 {c_2}+12 {c_4}+{c_6} \right) }{64 {\pi }^2 {s_1} {s_3}}
 {{\zeta _\eta}(3)}^2
-\frac{3 {c_1} \left( 3+2 {c_2} \right)}{8 {\pi }^3 {s_1}}
  {\zeta _\eta}(3) {\zeta'_\eta}(3)
\nonumber\\&&
-\frac{{{c_1}}^2 \left( 14+{c_2} \right) }{48 {\pi }^4}
 {{\zeta'_\eta}(3)}^2
\\
&& F\left[
\begin{array}{cccc}
+&-&-&+\\
-&+&+&-
\end{array}
\right]
\nonumber\\
&=&
 \frac{1}{4 \pi {c_1} {s_3}}
  {\zeta _\eta}(1)
+\frac{1}{12 {\pi }^2}
  {\zeta'_\eta}(1)
-\frac{ 5+18 {c_2}+7 {c_4}+{c_6} }{8 \pi {c_1} {s_3}}
  {\zeta _\eta}(3)
\nonumber\\&&
-\frac{ 1+6 {c_2} }{24 {\pi }^2}
  {\zeta'_\eta}(3)
-\frac{5 \left( 8+15 {c_2}+6 {c_4}+{c_6} \right)}{16 \pi {c_1} {s_3}}
  {\zeta _\eta}(5)
-\frac{ 2+3 {c_2} }{8 {\pi }^2}
  {\zeta'_\eta}(5)
\nonumber\\&&
+\frac{ 55+60 {c_2}+33 {c_4}+2 {c_6} }{16 {\pi }^2 {s_1} {s_3}}
  {\zeta _\eta}(1) {\zeta _\eta}(3)
+\frac{5 {c_1} \left( 3+2 {c_2} \right)  }{8 {\pi }^3 {s_1}}
  {\zeta'_\eta}(1) {\zeta _\eta}(3)
\nonumber\\&&
+\frac{{c_1} \left( 3+2 {c_2} \right)}{8 {\pi }^3 {s_1}}
  {\zeta _\eta}(1) {\zeta'_\eta}(3)
+\frac{{{c_1}}^2 \left( 14+{c_2} \right)}{24 {\pi }^4}
  {\zeta'_\eta}(1) {\zeta'_\eta}(3)
\nonumber\\&&
+\frac{15 \left( 20+27 {c_2}+12 {c_4}+{c_6} \right) }{32 {\pi }^2 {s_1} {s_3}}
  {\zeta _\eta}(1) {\zeta _\eta}(5)
+\frac{15 {c_1} \left( 3+2 {c_2} \right)  }{8 {\pi }^3 {s_1}}
  {\zeta'_\eta}(1) {\zeta _\eta}(5)
\nonumber\\&&
+\frac{3 {c_1} \left( 3+2 {c_2} \right)}{8 {\pi }^3 {s_1}}
  {\zeta _\eta}(1) {\zeta'_\eta}(5)
+\frac{{{c_1}}^2 \left( 14+{c_2} \right)}{8 {\pi }^4}
  {\zeta'_\eta}(1) {\zeta'_\eta}(5)
\nonumber\\&&
-\frac{9 \left( 20+27 {c_2}+12 {c_4}+{c_6} \right)}{64 {\pi }^2 {s_1} {s_3}}
 {{\zeta _\eta}(3)}^2
-\frac{3 {c_1} \left( 3+2 {c_2} \right)}{8 {\pi }^3 {s_1}}
  {\zeta _\eta}(3) {\zeta'_\eta}(3)
\nonumber\\&&
-\frac{{{c_1}}^2 \left( 14+{c_2} \right) }{48 {\pi }^4}
 {{\zeta'_\eta}(3)}^2
\end{eqnarray}
\end{document}